\documentclass[aps,pre,superscriptaddress] {revtex4}
\usepackage{amsmath}
\usepackage{bm}
\usepackage{graphicx}

\begin{document}

\title{Wandering breathers and self-trapping in weakly coupled nonlinear chains:
classical counterpart of macroscopic tunneling quantum dynamics}

\author{Yu.\ A.\ Kosevich}
\email[]{yukosevich@yahoo.com} \affiliation{Semenov Institute of
Chemical Physics, Russian Academy of Sciences, ul. Kosygina 4,
119991 Moscow, Russia}
\author{L. I. Manevitch}
\email[]{lmanev@center.chph.ras.ru} \affiliation{Semenov Institute
of Chemical Physics, Russian Academy of Sciences, ul. Kosygina 4,
119991 Moscow, Russia}
\author{A.V. Savin}
\email[]{asavin@center.chph.ras.ru} \affiliation{Semenov Institute
of Chemical Physics, Russian Academy of Sciences, ul. Kosygina 4,
119991 Moscow, Russia}

\begin{abstract}
We present analytical and numerical studies of phase-coherent
dynamics of intrinsically localized excitations (breathers) in a
system of two weakly coupled nonlinear oscillator chains. We show
that there are two qualitatively different dynamical regimes of
the coupled breathers, either immovable or slowly-moving: the
periodic transverse translation (wandering) of low-amplitude
breather between the chains, and the one-chain-localization of
high-amplitude breather. These two modes of coupled nonlinear
excitations, which involve large number of anharmonic oscillators,
can be mapped onto two solutions of a single pendulum equation,
detached by a separatrix mode. We also show that these two regimes
of coupled phase-coherent breathers are similar, and are described
by a similar pair of equations, to the two regimes in nonlinear
tunneling dynamics of two weakly linked Bose-Einstein condensates.
On the basis of this profound analogy, we predict a new tunneling
mode of two weakly coupled Bose-Einstein condensates in which
their relative phase oscillates around $\frac{\pi}{2}$ modulo
$\pi$. We also study two-chain breathers, which
can be considered as bound states of discrete breathers with
different symmetry and center locations in the coupled chains, and
bifurcation of the anti-phase two-chain breather into the
one-chain one. We show that static displacements of the coupled
chains with nonlinear localized excitation, induced by cubic term
in the intra-chain anharmonic potential, scale approximately as
the total vibration energy of the excitation, either one- or
two-chain one, and do not depend on the inter-chain coupling. Delocalizing transition of 1D
breather in 2D system of a large number of parallel coupled
nonlinear chains is described, in which the breather, initially
excited in a given chain, abruptly spreads its vibration energy in
the whole 2D system upon decreasing breather frequency or
amplitude below the threshold one. The threshold breather
frequency is above the cut off phonon frequency in 2D system, and
the threshold breather amplitude scales as square root of the
inter-chain coupling constant. Delocalizing transition of discrete
vibrational breather in 2D and 3D systems of coupled nonlinear
chains has an analogy with delocalizing transition for
Bose-Einstein condensates in 2D and 3D optical lattices.
\end{abstract}

\maketitle

\section{Introduction}

Nonlinear excitations (solitons, kink-solitons, intrinsically
localized modes and breathers) can be created most easily in
low-dimensional (1D and quasi-1D) systems
\cite{zabusky,koskov,dolgov,siev,page,kos1,kos2,sievpage,aubry,flach}.
Recent experiments have demonstrated the existence of
intrinsically  localized modes and breathers in various systems
such as 2D and 3D arrays of nonlinear optical waveguides
\cite{eisen,fleischer}, low-dimensional crystals \cite{swans},
antiferromagnetic materials \cite{sato}, Josephson junction arrays
\cite{trias}, photonic structures and micromechanical systems
\cite{campbell}, protein $\alpha$-helices \cite{edler}, and
$\alpha$-uranium \cite{manley}. Slowly-moving breathers and
supersonic kink-solitons were also described in 1D nonlinear
chains \cite{kos1,kos2,chen,kos3,kos4,kos5}, DNA macromolecules
\cite{yakush} and quasi-1D polymer crystals
\cite{savmanlam,avszub}.

One-dimensional arrays of magnetic or optical microtraps for
Bose-Einstein condensates (BECs) of ultracold quantum gases with
tunneling coupling provide a new field for the studies of coherent
nonlinear dynamics in low-dimensional systems
\cite{anderkas,kinosh}. In the mean-field theory, the tunneling
coupling between two BECs is similar to the linear coupling
between two optical waveguides \cite{corney} or between two
oscillator chains (nonlinear phononic waveguides). Here we show
that $phase$-$coherent$ dynamics of macroscopic ensembles of
classical particles (weakly-localized breathers) in two weakly
linked nonlinear oscillator chains has a profound analogy, and is
described by a similar in every respect pair of equations, to the
tunneling quantum dynamics of two weakly linked BECs in a
macroscopic double-well potential (single bosonic Josephson
junction) \cite{albiez}. The exchange of energy and excitations
between the coupled classical oscillator chains takes on the role
which the exchange of atoms via quantum tunneling plays in the
case of coupled BECs. Therefore such phase-coherent energy and
excitation exchange can be considered as a classical counterpart
of macroscopic tunneling quantum dynamics.

We show that there are two qualitatively different dynamical
regimes of the coupled breathers, the oscillatory exchange of the
low-amplitude breather between the chains ($wandering$
$breather$), and one-chain-localization (nonlinear self-trapping)
of the high-amplitude breather. These two regimes, which are
detached by a separatrix mode with zero rate of energy and
excitation exchange, are analogous to the two regimes in nonlinear
dynamics of macroscopic condensates in a single bosonic Josephson
junction \cite{albiez}. Essentially,  the phase-coherent  dynamics
of the coupled classical breathers is described by a pair of
equations, which coincides with the pair of coupled mean-field
equations describing coherent atomic tunneling in a single bosonic
tunnel junction \cite{smerzi,raghavan}. The predicted evolution of
the relative phase of two weakly coupled coherent breathers in
both dynamical regimes is also analogous to the evolution of
relative quantum mechanical phase between two weakly coupled
macroscopic condensates,  which was directly measured  in a single
bosonic Josephson junction by means of interference \cite{albiez}.
Moreover, the predicted  separatrix in the excitation exchange
between macroscopic ensembles of phase-coherent particles (weakly
localized breathers) in coupled oscillator chains with
``repulsive'' nonlinearity, which is determined by the ratio of
the intra-chain nonlinearity  (intra-chain interaction) and the
inter-chain coupling, can be considered as a classical nonlinear
dynamical model of the reversible interaction-induced
superfluid-Mott-insulator quantum phase transition, observed in
BEC with repulsive inter-atomic interaction in a lattice with
tunneling inter-site coupling \cite{bloch}. All these results
bring to light  a striking similarity, both in display and
evolution equations, between the classical phase-coherent
excitation exchange and macroscopic tunneling quantum dynamics
which can motivate new predictions and experiments in both fields.
On the basis of this profound analogy, we predict a new tunneling
regime of two weakly coupled BECs in which  their relative phase
oscillates around $\frac{\pi}{2}$ modulo $\pi$, which can be
observed by means of interference. This regime is different from
the regime of Josephson plasma oscillations, already realized in
experiments \cite{albiez}, when the relative phase of two weakly
linked BECs oscillates (or fluctuates \cite{gati}) around zero.

We also show that static displacements of the coupled chains with
nonlinear localized excitation, induced by cubic term in the
intra-chain anharmonic potential, scale approximately as the total
vibration energy of the excitation, either one- or two-chain one,
and do not depend on the inter-chain coupling. This observation is
also valid for a narrow stripe of several parallel coupled
nonlinear chains. We also study two-chain breathers which can be
considered as bound states of discrete breathers with different
symmetry and center locations in the coupled chains, and
bifurcation of the anti-phase  two-chain breather into the
one-chain one.  Bound states of two breathers with different
commensurate frequencies are also found in the two-chain system.
Merging of two breathers with different frequencies in one
breather in two coupled chains is described. Periodic transverse
translation (wandering) of low-amplitude breather in a system of
several, up to five, coupled nonlinear chains is observed, and the
dependence of the wandering period on the number of chains is
analytically estimated and compared with numerical results.
Delocalizing transition of 1D breather in 2D system of a large
number of parallel coupled nonlinear chains is described, in which
the breather, initially excited in a given 1D chain, abruptly
spreads its vibration energy in the whole 2D system upon
decreasing breather frequency or amplitude below the threshold
one. The threshold breather frequency is above the cut off phonon
frequency in 2D system, the threshold breather amplitude scales as
square root of the inter-chain coupling constant, and breather
vibration energy is localized mainly in one chain at the
delocalization threshold. Similar delocalizing transition for 1D
breather should occur also in 3D array of parallel coupled
nonlinear chains. Delocalizing transition of discrete breather in
2D and 3D systems of coupled nonlinear chains has an analogy with
delocalizing transition for polarons in 2D and 3D lattices
\cite{kalosak} and for Bose-Einstein condensates in 2D and 3D
optical lattices \cite{kalosak2}.

The paper is organized as follows. In the next Section II we
describe our model and the analytical predictions, derived on the
basis of this model. In Section III we describe our numerical
simulations and comparison with the analytical predictions, which
include simulation of dynamics of discrete breathers in two weakly
coupled chains (Section III A), simulation of
wandering breather in two weakly coupled chains (Section III B),
and simulation of breathers in a system of $M>2$ parallel weakly
coupled anharmonic chains (Section III C). In Section IV we give a
Summary of the main new results in this paper.

\section{Model and analytical predictions}

We consider two linearly coupled nonlinear oscillator  chains
(with unit lattice period). We model the chains with the
Fermi-Pasta-Ulam (FPU) Hamiltonian, which is one of the most
simple and universal models of nonlinear lattices and which can be
applied to a diverse range of physical problems \cite{chaos}:
\begin{eqnarray}
\label{e.1}
 H &=&
\sum_{i=1}^{2}\sum_{n}[\frac{1}{2}p_{n}^{(i)2}+\frac{1}{2}l^{(i)}
(u_{n+1}^{(i)}-u_{n}^{(i)})^2 + \frac{1}{3}\alpha^{(i)}(u_{n+1}^{(i)}-u_{n}^{(i)})^3 \nonumber \\
&& +\frac{1}{4}
\beta^{(i)}(u_{n+1}^{(i)}-u_{n}^{(i)})^4 + \frac{1}{2}C(u_{n}^{(i)}-u_{n}^{(3-i)})^2 ],
\end{eqnarray}
Here $u_{n}^{(i)}$ is displacement of the $n$-th particle from its
equilibrium position in the $i$-th chain,
$p_{n}^{(i)}=\dot{u}_{n}^{(i)}$ is particle momentum, $l^{(i)}$,
$\alpha^{(i)}$, $\beta^{(i)}$ and $C$ are, respectively,
intra-chain linear, intra-chain nonlinear and inter-chain linear
force constants. We assume that the coupling is weak,
$C$$\ll$$l^{(i)}$, and do not include the nonlinear inter-chain
interaction. The $\beta$-FPU Hamiltonian (\ref{e.1}) (with
$\alpha^{(i)}=0$) describes, in particular, purely transverse
particle motion \cite{kos1}. Torsion dynamics of DNA double helix
can also be approximated by the $\beta$-FPU Hamiltonian (1)
\cite{yakush}. On the other hand, weakly coupled nonlinear
molecular chains in polymers are characterized by the asymmetric
intra-chain anharmonic potential, with nonzero $\alpha^{(i)}$
\cite{savmanlam,avszub}.

Hamiltonian (1) generates the following equations of motion,
$i=1,2$:
\begin{eqnarray}
\label{e.2}
\ddot{u}_n^{(i)} &=&l^{(i)}(u_{n+1}^{(i)} + u_{n-1}^{(i)} - 2u_{n+1}^{(i)}) + \alpha^{(i)} [(u_{n+1}^{(i)} - u_{n}^{(i)})^2 -
(u_{n-1}^{(i)} - u_{n}^{(i)} )^2] \nonumber \\
 && + \beta^{(i)} [(u_{n+1}^{(i)} - u_{n}^{(i)})^3 + (u_{n-1}^{(i)} - u_{n}^{(i)})^3]+ C(u_{n}^{(3-i)}- u_{n}^{(i)}).
\end{eqnarray}

Below we consider two chains with identical anharmonic force
constants, when $\alpha^{(1)}=\alpha^{(2)}\equiv\alpha$ and
$\beta^{(1)}=\beta^{(2)}\equiv\beta$. We are interested in
high-frequency and therefore short-wavelength dynamics of the
coupled chains, when the displacements of the nearest-neighbor
particles are mainly anti-phase. For this case we can introduce
continuous envelope-functions for the particle displacements in
the chains, $f_{n}^{(i)}=u_{n}^{(i)}(-1)^n$, $f_{n}^{(i)}\equiv
f(x)_{i}$. These envelope-functions $f(x)_{i}$ are supposed to be
slowly varying on the interatomic scale in both chains, $\partial
f_{i}/\partial x <<1$, which allows us to write corresponding
partial differential equations for these functions, see, e.g.,
Refs. \cite{koskov,kos1,kos2,kos3}. To obtain nonlinear
envelope-function equations for the coupled chains  with
asymmetric inter-particle potential, one needs to consider
separately the dynamics of the $relative$,
$v_{n}^{(i)}=u_{n+1}^{(i)} - u_{n}^{(i)}$, and $total$,
$w_{n}^{(i)}=u_{n+1}^{(i)}+u_{n}^{(i)}$, displacements of the
nearest-neighbor particles,  cf. Ref. \cite{koskov}. Equations for
$v_{n}^{(i)}$ and $w_{n}^{(i)}$ can be obtained from Eqs.
(\ref{e.2}) and look as following:
\begin{eqnarray}
\label{e.3}
\ddot{v}_n^{(i)} &=&l^{(i)}(v_{n+1}^{(i)} + v_{n-1}^{(i)} - 2v_{n+1}^{(i)}) + \alpha [v_{n+1}^{(i)2} + v_{n-1}^{(i)2} - 2v_{n}^{(i)2}] \nonumber \\
&& +
\beta [(v_{n+1}^{(i)3} + v_{n-1}^{(i)3} - 2v_{n}^{(i)3}]+ C(v_{n}^{(3-i)}- v_{n}^{(i)}), \\
\ddot{w}_n^{(i)} &=& l^{(i)}(w_{n+1}^{(i)} + w_{n-1}^{(i)} -
2w_{n+1}^{(i)}) + \alpha [v_{n+1}^{(i)2} - v_{n-1}^{(i)2}]
\nonumber \\ &&  + \beta [v_{n+1}^{(i)3} - v_{n-1}^{(i)3}]+
C(w_{n}^{(3-i)}- w_{n}^{(i)}). \label{e.4}
\end{eqnarray}

Introducing continuous relative and total displacements,
$v_{n}^{(i)}\equiv v(x)_{i}$ and $w_{n}^{(i)}\equiv w(x)_{i}$, and
making in Eqs. (\ref{e.3}) and (\ref{e.4}) expansions of the
differences $u_{n\pm 1}^{(i)} - u_{n}^{(i)}$ and $w_{n\pm 1}^{(i)}
- w_{n}^{(i)}$ up to the second order, we get the following
partial differential equations for $v(x,t)_{i}$ and $w(x,t)_{i}$:
\begin{eqnarray}
\label{e.5} \ddot{v}_{i} &=&-l^{(i)}\frac{\partial^2
v_{i}}{\partial x^2} -4l^{(i)}v_{i} -4\frac{\partial
w_{i}}{\partial x}(\alpha v_{i}+\beta v_{i}^{2}) -4\beta v_{i}^{3}
+ C(v_{3-i} -v_{i}), \\
\ddot{w}_{i} &=&l^{(i)}\frac{\partial^2 w_{i}}{\partial x^2}
+2\alpha\frac{\partial v_{i}^{2}}{\partial x}
+2\beta\frac{\partial v_{i}^{3}}{\partial x} + C(w_{3-i} -w_{i}).
\label{e.6}
\end{eqnarray}

The main small parameters, which will allow us to find asymptotic
solutions of these equations, are the amplitudes of the
displacements $u_{n}^{(i)}$ of the coupled oscillations and hence
their envelopes $f(x,t)_{i}$. As follows from Eqs. (\ref{e.6}), in
the chains with asymmetric inter-particle potential (with finite
$\alpha$) the total displacements $w_{i}$ of the nearest neighbors
are weakly coupled (via anharmonic interaction) with relative
displacements $v_{i}$ and are in general small in comparison with
them, $w_{i} << v_{i}$ (see Eqs. (9)-(11) below). In such chains,
the asymmetric potential induces both the static (zero-harmonic)
and second-harmonic terms in the total displacements of the
nearest-neighbor particles, but the latter terms are smaller than
the former ones in the small-amplitude limit, cf. Ref. [2].

Now we analyze the static displacements $w_{0i}(x)$ of two
$identical$ weakly coupled chains with $l^{(1)}=l^{(2)}=1$. In
order to minimize the static elastic energy, the static
displacements of the chains should  be equal $at$ $both$
$infinites$: $w_{01}(x)=w_{02}(x)$ for $x\rightarrow -\infty$ and
$x\rightarrow +\infty$. To find $w_{0i}(x)$, we take the sum and
the difference of Eqs. (\ref{e.6}) for the two chains to obtain
the following equations:
\begin{eqnarray}
\label{e.7}
\ddot{w}_{1}+\ddot{w}_{2}&=&\frac{\partial^2 (w_{1}+ w_{2})}{\partial x^2} +2\alpha\frac{\partial (v_{1}^{2}+v_{2}^{2})}{\partial x}
+2\beta\frac{\partial (v_{1}^{3}+v_{2}^{3})}{\partial x}, \\
\ddot{w}_{1}-\ddot{w}_{2}&=&\frac{\partial^2 (w_{1}-
w_{2})}{\partial x^2} - 2C(w_{1}- w_{2}) +2\alpha\frac{\partial
(v_{1}^{2}- v_{2}^{2})}{\partial x} +2\beta\frac{\partial
(v_{1}^{3}- v_{2}^{3})}{\partial x}. \label{e.8}
\end{eqnarray}

Omitting the time-derivative terms, we get from Eq. (\ref{e.7})
the following expression for the static center-of-mass
displacement of the two chains:
\begin{equation}
\label{e.9} w_{0}^{(CM)}(x)=\frac{w_{01}(x)+
w_{02}(x)}{2}=-\alpha\int_{-\infty}^{x} \langle
v_{1}^{2}+v_{2}^{2}\rangle  dx^{'},
\end{equation}
where the angle brackets denote the time averaging. The nonlinear
term, proportional to $\beta\langle v_{1}^{3}+v_{2}^{3}\rangle$,
 cf. Eq. (\ref{e.7}),
 is averaged out in Eq. (9) because of the third power of $v^{(i)}$.
  At both
infinities, the displacement $w_{0}^{(CM)}(x)$ coincides with
equal static displacements of the coupled chains,
$w_{0}^{(CM)}(\pm\infty)=w_{0}^{(1)}(\pm\infty)=w_{0}^{(2)}(\pm\infty)$.
It is assumed in Eq. (9) that $w_{0}^{(CM)}(-\infty)=0$,
$v^{(i)}(-\infty)=0$ and $\frac{\partial w_{0}^{(CM)}}{\partial x}
(-\infty) =0$, $i=1,2$. Therefore the static displacements of the
both coupled chains $w_{0}(x)^{(i)}$ have in general the form of
the  $kink$-$like$ pattern with equal center-of-mass displacement
difference $\Delta
w_{0}^{(CM)}=w_{0}^{(CM)}(+\infty)-w_{0}^{(CM)}(-\infty)=w_{0}^{(CM)}(+\infty)$.
The kink-like static displacement pattern of intrinsically
localized mode (breather) in a $single$ $\alpha$-$\beta$-FPU chain
was studied in Refs. \cite{koskov,bickham,huang,kiselev}.

The term in angle brackets in r.h.s. of Eq. (9) coincides with the
total vibration energy of two coupled chains (in the low-amplitude
limit when the energy  is governed mainly by the quadratic terms
in Hamiltonian (1)). Therefore the center-of-mass displacement of
the coupled chains with asymmetric anharmonic potential scales
approximately as the $total$ $vibration$ $energy$  of the
nonlinear localized excitation, the one- or two-chain breather in
two coupled chains: $w_{0}^{(CM)}=-\alpha\int_{-\infty}^{+\infty}
\langle v_{1}^{2}+v_{2}^{2}\rangle  dx^{'}$. Equations (7) an (9)
also show that static $strain$ of the chains, given by the
displacement derivative, is determined by the $local$ $density$ of
the total energy of the nonlinear localized excitation: $\partial
w_{0}^{(CM)}(x)/\partial x =-\alpha\langle
v_{1}^{2}+v_{2}^{2}\rangle(x)$. These conclusions are confirmed by
our numerical simulations, see Figs. 2(c) and 2(d) below.

These conclusions can be easily generalized to the system of $M>2$
parallel coupled nonlinear chains, see also Section III C below.
In this case only the last term in Eqs. (5) and (6), describing
the inter-chain coupling, will be modified (cf. Eq. (62) below).
In a narrow stripe of chains, when the number of particles $N$ in
each chain greatly  exceeds the number of chains, $N\gg M$, the
static center-of-mass displacement in the multi-chain system will
have the form:
\begin{equation}
\label{e.9}
w_{0}^{(CM)}(x)=\frac{1}{M}\sum_{m=1}^{M}w_{0m}(x)=-\frac{2\alpha}{M}\int_{-\infty}^{x}
\langle \sum_{m=1}^{M}v_{m}^{2}\rangle dx^{'},
\end{equation}
where $v_{m}(x)$ and $w_{0m}(x)$ are, respectively, the relative
and static total nearest-neighbor displacements in the $m$-th
chain. It is assumed in Eq. (10) that $w_{0}^{(CM)}(-\infty)=0$,
$v_{m}(-\infty)=0$ and $\frac{\partial w_{0}^{(CM)}}{\partial x}
(-\infty)=0$. As follows from Eq. (10), the lattice displacement
$w_{0}^{(CM)}(x)$ and the corresponding lattice strain $\partial
w_{0}^{(CM)}/\partial x$ in the multi-chain system depend on the
number of the excited chains and their excitation energies.
Importantly, the static lattice displacements and strain in a
system of coupled chains, Eqs. (9) and (10), do not depend on the
strength of the inter-chain coupling, which is connected with
long-range action of static elastic fields.

The magnitudes of the static displacements in the central region
of the kinks in the two chains are different in general and can be
found with the help of Eq. (8) and the above requirements for the
values of $w_{0}(x)^{(i)}$ at the infinities. Omitting the
time-derivative terms, we get from Eq. (\ref{e.8}) the following
solution for the difference of static displacements of the two
coupled chains:
\begin{eqnarray}
\Delta w_{0}(x)&\equiv &w_{01}(x)-
w_{02}(x)=\frac{1}{4}\sqrt{\frac{2}{C}}\langle\int_{-\infty}^{x}\sinh[(s-x)\sqrt{2C}]A(s)ds+
\nonumber \\ & &
+\int_{x}^{\infty}\sinh[(x-s)\sqrt{2C}]A(s)ds-\int_{-\infty}^{-x}\sinh[(s+x)\sqrt{2C}]A(s)ds+\nonumber
\\ & & +\int_{-x}^{\infty}\sinh[(s+x)\sqrt{2C}]A(s)ds \rangle.
\end{eqnarray}
where $A(s)=\alpha\partial (v_{1}^{2}(s) - v_{2}^{2}(s))/\partial
s$. Taking into account that $v_{i}^{2}(x)$ is a symmetric
function with respect to the breather center $x=0$, see Eqs. (20)
and (21) below, Eq. (11) gives us that $\Delta w_{0}(x)=-\Delta
w_{0}(-x)$ and therefore $\Delta w_{0}(x)$ is zero both at $x=0$
and at both infinities. [As follows from Eq. (11), the difference
of static displacements of the coupled chains $\Delta w_{0}(x)$
depends in general on the coupling constant $C$, in contrast to
the static center-of-mass displacement $w_{0}^{(CM)}(x)$, Eq.
(9).] As follows in turn from Eqs. (6), the second-harmonic terms
in the total displacements $w(x)^{i}$ in the chains have the
smallness of the third order (in $v(x)^{(i)}$) because of the
smallness of spatial derivative $\partial v^{(i)2}/\partial x \sim
\lambda_{i}v^{(i)2} \sim v^{(i)3}$, see Eqs. (19), (20) and (23)
below. This allows us to neglect in the following analysis of
breather dynamics in coupled chains both the difference of static
displacements of the chains $\Delta w_{0}(x)$ and the
second-harmonic terms in the total displacements $w(x)^{i}$ in
comparison with the static center-of-mass displacement
$w_{0}^{(CM)}(x)$.

Taking into account that $v(x)_{i}=2f(x)_{i}$, we get from  Eqs.
(5), (7) and (9) the following coupled nonlinear partial
differential equations for the real envelope-functions $f(x)_{i}$,
$i=1,2$:
\begin{equation}
\ddot{f}_{i}+\omega_{mi}^{2}f_{i} + \frac{\partial^2
f_{i}}{\partial x^2}+16\beta f_{i}^{3} - 16\alpha^{2}\langle
f_{1}^{2}+f_{2}^{2}\textsc{}\rangle f_{i} -Cf_{3-i}=0,
\end{equation}
where $\omega_{mi} =\sqrt{4l^{(i)} +C}$ is characteristic
frequency slightly above the maximal phonon frequency in the
$i$-the isolated chain. These self-consistent equations describe
the dynamical relative lattice displacements $v_{i}(x)$, while the
static lattice displacements $w_{0i}(x)$ can be obtained with the
use of Eqs. (9) and (11).  It is worth mentioning that collective
static deformation of the coupled chains, induced by the
asymmetric intra-chain anharmonic potential, makes the coupled
envelope-function equations (12) $non$-$local$ in general, via the
term proportional to $-\alpha^2$.

In order to deal with the amplitude and phase of the coupled
nonlinear excitations, it is useful to introduce complex wave
fields $\Psi(x,t)_i$ for each chain, see, e.g., Ref. \cite{kos3}:
\begin{equation}
f(x,t)_{i} = \frac{1}{2}\left[\Psi(x,t)_{i}+
\Psi(x,t)_{i}^{\ast}\right].
\end{equation}

Then in resonance approximation we get from Eqs. (12) and (13) the
following coupled equations for $\Psi(x,t)_i$, $i=1,2$:
\begin{equation}
\frac{1}{2}(\frac{\partial^2 \Psi_{i}}{\partial t^2} +
\frac{\partial^2 \Psi_{i}}{\partial x^2} + \omega_{mi}^2\Psi_{i})+
6\beta\mid\Psi_{i}\mid^2 \Psi_{i} - 8\alpha^2 \langle
\mid\Psi_{1}\mid^2 +
\mid\Psi_{2}\mid^2\rangle\Psi_{i}-\frac{1}{2}C\Psi_{3-i}=0,
\end{equation}
(and complex-conjugated equations for $\Psi_{i}^{\ast}$). This
approximation, in which we neglect the higher harmonics, is valid
for the dispersive system under consideration due to the weakness
of nonresonant interaction between the mode with fundamental
frequency and its third harmonic, cf. Ref. \cite{kos3}.

Using Eqs. (14) in the assumption that the shift of the wandering
breather frequency, caused by nonlinearity, weak inter-chain
coupling and slow breather motion along the chains, is relatively
small, one can readily show the existence in general of the
following three integrals of motion:
\begin{eqnarray}
N_b &=& \int \left[\mid\Psi_1\mid^2 + \mid\Psi_2\mid^2 \right]dx, \\
E_b &=& \int[\sum_{i=1}^{2}(3\beta\mid\Psi_i\mid^4 -8\alpha^2 \langle \mid\Psi_{1}\mid^2 + \mid\Psi_{2}\mid^2\rangle \mid\Psi_{i}\mid^2 -\nonumber \\ && -\frac{1}{2}\mid\frac{\partial\Psi_i}{\partial x}\mid^2 )-\frac{1}{2}C(\Psi_1\Psi_2^\ast + \Psi_2 \Psi_1^\ast)]dx, \\
P_{bx} &=& -\frac{i}{2}\int\sum_{i=1}^{2}\left[\Psi_i
\frac{\partial\Psi_i^\ast}{\partial x} - \Psi_i^\ast
\frac{\partial\Psi_i}{\partial x}\right]dx,
\end{eqnarray}
which play the role of breather total number of excitations, total
energy and total momentum along the chain axis, respectively, cf.
\cite{kos3}. The existence of these integrals of motion
demonstrates that the exchange of energy and momentum between two
nonlinear  systems (1) is a coherent phenomenon, which depends in
general on the initial  excitation conditions.

It is also useful to define  partial numbers of excitations
\begin{equation*}
 N_{i} = \int \mid\Psi_{i}\mid^2dx,~~~~~i=1,2,
\end{equation*}
when $\dot{N}_{i}+J_{i}=0$, where
\begin{equation}
 J_{i} = -J_{3-i} = \frac{iC}{2\omega}\int \left[\Psi_{i}\Psi_{3-i}^{\ast} - \Psi_{i}^{\ast}\Psi_{3-i}\right]dx
\end{equation}
 is a total inter-chain flux of excitations, which conserves the total number of them: $\dot{N}_{b}=0$
 where  $N_{b} = N_{1}+ N_{2}$,
 $\omega$ is frequency of wandering breather, see Eqs. (19), (20) and (22) below.

To describe with the help of Eqs. (14) a slowly-moving breather,
wandering between two weakly coupled nonlinear chains with
positive (repulsive) anharmonic force constant $\beta$, we assume
the following form for the complex fields $\Psi_1$ and $\Psi_2$:
\begin{eqnarray}
\Psi_1 &=&\Psi_{max}\frac{\exp[i(kx-\omega
t)]}{\cosh[\lambda_{1}(x-Vt)]}\cos\Theta
\exp(-\frac{i}{2}\Phi), \\
\Psi_2 &=&\Psi_{max}\frac{\exp[i(kx-\omega
t)]}{\cosh[\lambda_{2}(x-Vt)]}\sin\Theta \exp(\frac{i}{2}\Phi),
\end{eqnarray}
where $\omega$, $V\ll 1$ and $k\ll 1$ are, respectively,
frequency, slow velocity and small wavenumber related with the
moving breather, $\lambda_{1,2}\ll 1$ are real parameters
describing inverse localization lengths of the breathers;
$\Phi=\Phi(t-kx/\omega)$ stands for the relative phase of the
coupled chains, while  the parameter $\Theta=\Theta(t-kx/\omega)$
describes the "relative population" of the two chains
$z=(n_{1}-n_{2})/(n_{1}+n_{2})=\cos2\Theta$, where
$n_{i}$$=$$\mid$$\Psi_{i}$$\mid^2$ is local density of excitations
in the $i$-th chain, and $\langle \mid\Psi_{1}\mid^2 +
\mid\Psi_{2}\mid^2\rangle=\Psi_{max}^2$ in Eq. (14). The
parameters $\Theta$ and $\Phi$ determine also the inter-chain flux
of excitations, cf. Eq. (18):
\begin{equation} J_{1}=-J_{2}=\frac{C\Psi_{max}^2}{2\omega}\int\frac{\sin2\Theta\sin\Phi}{\cosh[\kappa_{1}
(x-Vt)]\cosh[\kappa_{2}(x-Vt)]}dx.
  \label{e.13}
\end{equation}

Using Eqs. (14), (19) and (20),
 after some algebra we obtain dispersion equations for the introduced
parameters,
\begin{eqnarray}
 \label{e.14}
 \omega^2&=&\frac{1}{2}(\omega_{m1}^2 +\omega_{m2}^2)+(3\beta -8\alpha^2)\Psi_{max}^2 - k^2 -C\frac{\cos\Phi}{\sin(2\Theta)}, \\
 \lambda_1^2&=&(6\beta\cos^2 \Theta - 8\alpha^2)\Psi_{max}^2,~~~ \lambda_2^2 = (6\beta\sin^2 \Theta  - 8\alpha^2)\Psi_{max}^2, \\
 V&=&\frac{\partial\omega}{\partial k},
  \label{e.16}
\end{eqnarray}

and evolution equations for $\Phi$ and $\Theta$:
\begin{eqnarray}
 \label{e.17}
\dot{\Phi}&=&\frac{1}{2\omega}(\omega_{m1}^2 -\omega_{m2}^2)
+\frac{3\beta\Psi_{max}^2}{\omega}\cos(2\Theta) +
\frac{C}{\omega}\cos\Phi\cot(2\Theta), \\
\dot{\Theta}&=&\frac{C}{2\omega}\sin\Phi.
 \label{e.18}
\end{eqnarray}

In the derivation of Eqs. (\ref{e.17}) and (\ref{e.18}), it was
assumed explicitly that the ratio
$\frac{\cosh[\lambda_{1}(x-Vt)]}{\cosh[\lambda_{2}(x-Vt)]}$ is
equal to one. The latter is valid for small-amplitude  breathers
with long localization lengths, $\lambda_{1,2}\ll 1$. In this case
the above assumption, which is exact for the central region of the
breathers, $x$-$Vt$$\approx 0$, will be (approximately) valid for
a large number of particles, which form weakly localized wandering
breather in weakly coupled nonlinear chains. It is also assumed in
the considered approximation that the shifts of breather frequency
$\omega$, Eq. (22), caused by weak inter-chain coupling $C$,
nonlinearity $(3\beta -8\alpha^2)\Psi_{max}^2$ and slow breather
motion along the chains $k^2$, as well as the characteristic
frequency difference $\mid$$\omega_{m1}$
-$\omega_{m2}$$\mid$/$\omega$, are all relatively small. It is
worth to mention that the parameter $-8\alpha^2 \Psi_{max}^2$
determines the shift of breather frequency $\omega$ and inverse
localization lengths $\lambda_{1,2}$, see Eqs. (22) and (23), but
does not enter evolution equations (25) and (26) for $\Theta$ and
$\Phi$. Equations similar to Eqs. (\ref{e.17}) and (\ref{e.18})
were probably derived for the first time
 in Ref. \cite{ovchin} for the description
of energy exchange between two weakly coupled classical anharmonic
oscillators (with $\omega_{m1}=\omega_{m2}$ and $\alpha=0$). Later
similar equations were used for the description of the power
exchange between two weakly coupled nonlinear optical waveguides
\cite {jensen} and of the dynamics of coupled nonlinear
oscillators \cite{koskov2}.

Equations (25) and (26) can be written in an equivalent form for
the relative phase $\Phi$ and population imbalance $z$, when
$z=\cos2\Theta$ and $\sqrt{1-z^2}=\sin2\Theta$:
\begin{eqnarray}
\label{e.19}
\dot{\Phi}&=&\omega_{m1}-\omega_{m2}+\frac{3\beta\Psi_{max}^2}{\omega}z
+ \frac{C}{\omega}\frac{z}{\sqrt{1-z^2}}\cos\Phi, \\
\dot{z}&=&-\frac{C}{\omega}\sqrt{1-z^2}\sin\Phi.  \label{e.20}
\end{eqnarray}

Here the variables  $\Phi$ and $z$ are canonically conjugate,
\begin{equation}
\dot{\Phi}=\frac{\partial H_{eff}}{\partial z}, ~~~~
\dot{z}=-\frac{\partial H_{eff}}{\partial \Phi},
\end{equation}
 with the following effective Hamiltonian (which has the dimension of
 frequency):
\begin{equation}
H_{eff}=\frac{3\beta\Psi_{max}^2}{2\omega}z^2-\frac{C}{\omega}\sqrt{1-z^2}\cos\Phi+z(\omega_{m1}-\omega_{m2}).
\end{equation}

The  very same equations (\ref{e.19}) and (\ref{e.20}) for $\Phi$
and $z$, which are equivalent to Eqs. (\ref{e.17}) and
(\ref{e.18}) for $\Phi$ and $\Theta$, were obtained in
\cite{smerzi,raghavan} in the mean-field theory of tunnelling
dynamics of two weakly coupled Bose-Einstein condensates, which
were later used in the analysis of the experimental realization of
a single bosonic Josephson junction \cite{albiez}.  Therefore
generic evolution equations (27) and (28) (being written in
corresponding units) for the dynamics of two coupled
phase-coherent breathers or tunneling dynamics of two coupled BECs
do not explicitly depend on the "source" equations: Eqs. (14) for
coupled oscillator chains or mean-field Gross-Pitaevskii equations
for coupled BECs \cite{smerzi,raghavan}. In our case, Eqs.
(\ref{e.19}) and (\ref{e.20}) describe the exchange of lattice
excitations between the chains rather than atomic tunneling. One
can therefore consider such excitation exchange as a classical
counterpart of tunneling quantum dynamics. Some other equivalent
forms of the effective Hamiltonian (30) were also discussed in
Refs. \cite{tsironis,christian}.

It is noteworthy that equations, similar to Eqs.\,(\ref{e.17}) and
(\ref{e.18}), describe the dynamics of two weakly coupled
nonlinear oscillators (with different harmonic eigenfrequencies
$\omega_{m1}$ and $\omega_{m2}$) \cite{koskov2,manev1}. Therefore
wandering breathers can be considered as weakly coupled
phase-coherent nonlinear $macroscopic$ $oscillators$.

Equations (25) and (26) can be solved analytically  for the given
initial conditions with the use of variational method. For two
$identical$ chains, with
$\omega_{m1}$=$\omega_{m2}$$\equiv$$\omega_{m}$, we assume that
the solution of Eq. (26) has the following form:
\begin{equation}
\cos\Phi=A(t)/\sin(2\Theta),
\end{equation}
where $A$=0 for $\sin(2\Theta)$=0. Using this ansatz, we get from
Eqs. (25) and (26) that
\begin{equation}
\dot{A}=
-6\frac{\beta\Psi_{max}^2}{C}\sin(2\Theta)\cos(2\Theta)\dot{\Theta}.
\end{equation}
For the aforementioned initial condition, when $A$=0 for
$\sin(2\Theta)$=0, from Eq. (32) we find:
\begin{eqnarray}
A&=&-\frac{3\beta\Psi_{max}^2}{2C}\sin^2 (2\Theta), \\
\cos\Phi&=&-\frac{3\beta\Psi_{max}^2}{2C}\sin(2\Theta)=-\frac{3\beta\Psi_{max}^2}{2C}\sqrt{1-z^2},
\label{e.21}
\end{eqnarray}
which corresponds to $\Phi=\pi/2$ modulo $\pi$  for
$\sin(2\Theta)=0$. This exact solution of Eqs. (\ref{e.17}) and
(\ref{e.18}), (\ref{e.19}) and (\ref{e.20}), conserves the
effective Hamiltonian:
$H_{eff}$=$\frac{3\beta\Psi_{max}^2}{2\omega}$. The important
feature of this solution is that the relative phase $\Phi$ is
$self$-$locked$ to the value $\frac{\pi}{2}$ modulo $\pi$ by the
total population imbalance $\mid z\mid$=1 of the two coupled
chains. The phase portrait of Eq. (\ref{e.21}) in the $\Phi - z$
plane is given by
\begin{equation}
(\kappa\cos\Phi)^2+z^2=1,
\end{equation}
 where
$\kappa=2C/3\beta\Psi_{max}^2$, see Fig. 1.

With the use of Eq. (\ref{e.21}), Eqs. (\ref{e.14}), (\ref{e.17})
and (\ref{e.18}), (\ref{e.19}) and (\ref{e.20}) will take the
following form:
\begin{eqnarray}
\omega^2&=&\omega_{m}^2+(\frac{9}{2}\beta -8\alpha^2)\Psi_{max}^2
- k^2, \\
\dot{\Phi}&=&\frac{3}{2\omega}\beta\Psi_{max}^2\cos(2\Theta),~~~\dot{\Theta}=\frac{C}{2\omega}\sin\Phi,\\
\dot{\Phi}&=&\frac{3}{2\omega}\beta\Psi_{max}^2z,
~~~\dot{z}=-\frac{C}{\omega}\sqrt{1-z^2}\sin\Phi.
\end{eqnarray}

Finally, for two identical weakly coupled chains with
$l^{(1)}=l^{(2)}=1$ and $\omega_m\approx 2$ we get from Eqs. (37)
the following two equivalent physical-pendulum equations:
\begin{equation}
\label{e.39} \ddot{\Xi}+\Omega_0^2\sin\Xi =0
\end{equation}
- for $\Xi=4\Theta$ where $\Omega_0=3\beta\Psi_{max}^2/4$, and
\begin{equation}
\ddot{\delta}+\frac{C^2}{4}\sin\delta =0
 \label{e.27}
\end{equation}
- for $\delta=2\Phi -\pi$. In the following we will solve Eq.
(\ref{e.39})  with the initial condition $\Theta(0)=0$, which
corresponds to zero complex field $\Psi_2$ in the second chain at
$t$$=$0 (or $z(0)$=1), and which is realized in our simulations.
Therefore  we will assume that $\Xi(0)=0$, $\Phi(0)=\pi/2$  modulo
$\pi$ and $\dot{\Xi}(0)=C$. The corresponding initial conditions
for $\delta$ in Eq. (40) are $\delta(0)=0$  modulo $2\pi$ and
$\dot{\delta}(0)=\frac{3}{2}\beta\Psi_{max}^2$.

Solution of the nonlinear physical-pendulum equation is well
known, see, e.g., Ref. \cite{lichtenberg}, and can be written in
terms of elliptic functions (with the elliptic modulus
$\kappa$=$2C/3\beta\Psi_{max}^2$). Namely, one can obtain from Eq.
(39) the following time evolution and fundamental oscillation
frequency $\omega_{4\Theta}$  for  $\Xi=4\Theta$:
\begin{eqnarray}
4\Theta&=&2\arcsin[\kappa sn(\Omega_{0}t,\kappa )],~~~~ \kappa \leq 1, \\
4\Theta&=&2\arcsin[sn(\kappa \Omega_{0}t,1/\kappa)],~~~~ \kappa \geq 1, \\
\omega_{4\Theta} &=&\frac{\pi}{2}\frac{\Omega_0}{K(\kappa )}, ~~~~\kappa \leq 1, \\
\omega_{4\Theta} &=&\pi\frac{\Omega_0 \kappa }{K(1/\kappa )}, ~~~~ \kappa
\geq 1,
\end{eqnarray}
where $sn$ is the Jacobian elliptic sine,
$K(\kappa)=F(\pi/2,\kappa)$ is complete elliptic integral of the
first kind. The solution for $\delta=2\Phi -\pi$ can also be found
with the use of Eq. (40).

The important property of the solution of the physical-pendulum
equation (39) (or (40)) is the existence of two qualitatively
different dynamical regimes of Eqs. (37) and (38), which are
detached by a separatrix corresponding to the condition $\kappa=1$
or $3\beta\Psi_{max}^2 =2C$.

For $\kappa$$\gg$1 or $\beta$$\Psi_{max}^2$$\ll$$2C/3$, the
parameter $\Theta$ linearly grows with the ``running'' time
$\tilde{t}$$\equiv$$t-\frac{k}{\omega}x$, while
$\Phi\approx\Phi(0)=\frac{\pi}{2}$ (see, e.g., Ref. \cite{kos6}):
\begin{equation}
\Theta\approx
\tilde{C}\frac{\tilde{t}}{4}+\frac{9\beta^2\Psi_{max}^4}{64\tilde{C}^{2}}\sin(\tilde{C}\tilde{t}),
~~\Phi\approx\frac{\pi}{2}
+\frac{3\beta\Psi_{max}^2}{2\tilde{C}}\sin(\tilde{C}\frac{\tilde{t}}{2}),
~~z\approx \cos(\tilde{C}\frac{\tilde{t}}{2}),~~
\tilde{C}=C-\frac{\Omega_{0}^{2}}{C}=C-\frac{9\beta^2\Psi_{max}^4}{16C}.
\label{e.28}
\end{equation}

In this regime $\Theta$ spans the full range from $0$ to $2\pi$,
which, according to Eqs. (19) and (20), corresponds to the
$complete$ energy exchange between the nonlinear chains and
therefore to the breather, wandering between the two chains.
According to Eq. (21), the flux of the inter-chain energy exchange
in this regime $J_1\propto
C\Psi_{max}\sin2\Theta\sin\Delta=C\Psi_{max}\sin(\tilde{C}\tilde{t}/2)$
has the rate,
$\omega_{beat}=\tilde{C}/2=(C-9\beta^2\Psi_{max}^4/(16C))/2$ for
$\kappa$$\gg$1, which continuously decreases with the increase of
the ratio $\beta\Psi_{max}^2/C$ below the separatrix. Since
$\left<\cos\Theta^2\right>=\left<\sin\Theta^2\right>=1/2$ in this
mode, the time-averaged inverse localization lengths
$\lambda_{1,2}$, Eq. (23), are equal in the two chains:
$\lambda_1=\lambda_2=\Psi_{max}\sqrt{3\beta - 8\alpha^2}$ (for
$3\beta > 8\alpha^2$, see Section III A below).

This dynamical regime is analogous to the complete energy exchange
in the beating in a system of two coupled harmonic oscillators.
The solution $\Theta =C\tilde{t}/4$, $\Phi =\frac{\pi}{2}$ and
$z=\cos(C\tilde{t}/2)$ can be obtained directly from the
linearized Eqs. (14).

Similar tunneling mode can also be realized for BEC in a symmetric
double-well potential when the condensate is initially loaded into
one of the wells, $z(0)$=$\pm 1$, cf. Ref. \cite{corney}. In such
mode the relative phase of the coupled BECs
will oscillate around $\frac{\pi}{2}$ modulo $\pi$, see
Eq.\,(\ref{e.28}) and Fig.1, which can be observed by means of
interference. This tunneling mode is similar to but is different
from the mode of Josephson plasma oscillations, already realized
in experiments \cite{albiez}, in which the relative population
imbalance is always less than one and the relative phase of the
coupled BECs oscillates around zero (modulo $2\pi$). The
observable difference between the time evolution of average
relative phases in the observed in \cite{albiez} tunneling mode
and the mode predicted in Fig. 1 is caused by the difference in
the initial states (population imbalance and relative phase) of
two weakly linked Bose-Einstein condensates: it is $\mid z(0)\mid
<1$ and $\Phi(0)=0$ modulo $2\pi$  in the former case while it is
$\mid z(0)\mid=1$ and $\Phi(0)=\pi/2$ modulo $\pi$ in the latter
case.

It is worth to emphasize that according to Eqs. (19), (20) and
(45), the temporal Fourier spectrum of the wandering breather is
determined both by the frequency $\omega$ and time dependence of
$\Theta$. The latter produces in spectral density of the wandering
breather a series of doublets of frequencies, shifted upward and
downward with respect to the fundamental breather frequency
$\omega$: $\omega \pm n\omega_{beat}=\omega\pm n\tilde{C}/2$,
$n=1,2, ...$, see Eq. (45). The lower frequencies in such doublets
can become less than the threshold frequency $\omega_{m}$ in the
phonon band of the coupled chains, see the dashed line in  Fig. 8
below, which results in weak damping of the coupled nonlinear
oscillations caused by phonon emission in the chains.

In the opposite limit $k\ll 1$ or $\beta\Psi_{max}^2 \gg 2C/3$,
one has
\begin{eqnarray}
\Theta &\approx&
\frac{C}{3\beta\tilde{\Psi}_{max}^2}\sin(\frac{3}{4}\beta\tilde{\Psi}_{max}^2
\tilde{t}),~~\Phi\approx\frac{\pi}{2} +
\frac{3}{4}\beta\tilde{\Psi}_{max}^2
\tilde{t}+\frac{C^2}{18(\beta\tilde{\Psi}_{max}^2)^2}\sin(\frac{3}{2}\beta\tilde{\Psi}_{max}^2\tilde{t}), \nonumber \\
z&\approx &1 -
\frac{2C^2}{9(\beta\tilde{\Psi}_{max}^2)^2}\sin^{2}(\frac{3}{4}\beta\tilde{\Psi}_{max}^2
\tilde{t}),~~~
\beta\tilde{\Psi}_{max}^2=\beta\Psi_{max}^2-\frac{C^2}{9\beta\Psi_{max}^2}.
\end{eqnarray}

In this mode the inter-chain energy exchange between the coupled
chains is incomplete since always $\Theta\ll 1$, cf. Eqs. (19) and
(20). With the increase of the ratio $\beta\Psi_{max}^2 /C$ beyond
the separatrix, the flux of such incomplete energy exchange,
$J_1\propto (C^2/\Psi_{max})\sin2\Theta\sin\Phi
=(C^2/2\Psi_{max})\sin(3\beta\tilde{\Psi}_{max}^2 \tilde{t}/2)$,
cf. Eq. (21), gradually decreases but its rate
$3\beta\tilde{\Psi}_{max}^2/2$ gradually increases, see Eq. (46).
This mode is similar to the macroscopic quantum self-trapping of
Bose-Einstein condensate in a single bosonic Josephson junction
\cite{corney,smerzi,raghavan,albiez}, as well as to the asymmetric
nonlinear mode (known, e.g., for two coupled nonlinear waveguides
\cite{jensen,uzun,akhmediev}), in which one system, here is system
1, carries almost all vibrational energy while the other one is
almost at rest.

The separatrix  $k=1$ or $\beta\Psi_{max}^2 =2C/3$ is
characterized by zero frequency of the physical pendulum (39) (or
(40)), which corresponds to the infinite period of inter-chain
energy exchange. For the considered initial conditions, the
separatrix is described by the following solution of Eqs. (37) and
(38):
\begin{equation}
\Theta=\arctan[\exp(\frac{C}{2}t)]-\frac{\pi}{4},~~\Phi=2\arctan[\exp(\frac{C}{2}t)],
~~z(0)=1,~~
z(\infty)=0,~~\Phi(0)=\frac{\pi}{2},~~\Phi(\infty)=\pi.
\end{equation}

The flux of the inter-chain excitation exchange is zero at the
separatrix for $t\rightarrow\infty$: $J_{i}=0$ since $\sin\Phi=0$,
see Eq. (21). On the other hand, the breather frequency $\omega$
is finite  at the separatrix and is equal to
\begin{equation}
\omega_{sep} =\omega_m + (9\beta/8 -2\alpha^2)\Psi_{max}^2 -k^2
/4\approx 2+ C-2\alpha^2\Psi_{max}^2 -k^2 /4,
\end{equation}
since $3\beta\Psi_{max}^2=2C$ at the separatrix and
$\omega_m\approx 2+ C/4$, see Eqs. (12), (19) and (36).  The value
$\omega_{sep}\approx 2.1$ for $C=0.1$,  $\alpha=0$ and $k\ll 1$
coincides with good accuracy with the one, $\omega_{sep}\approx
2.098$, which follows from our numerical simulations of the
separatrix-like dynamics of the wandering breather in two coupled
$\beta$-FPU chains, see Figs. 7 and 11 below. In corresponding
dimensionless units, the above separatrix, Eq. (47), gives the
position, in terms of spatial soliton amplitude and inter-fiber
coupling, of the numerically revealed separatrix-like regime
between the regimes of the total and partial exchange of photon
energy between two coupled nonlinear optical fibers (fiber
directional coupler), described by two coupled nonlinear
Schr\"odinger equations \cite{1999}. The existence of a separatrix
was revealed in different models of nonlinear dimer (two coupled
nonlinear systems with two degrees of freedom), both classical and
quantum, in Refs. \cite{ovchin,jensen,koskov2,eilbeck,aubry2}.

The important result of our studies in this field is that the
dynamics of two coupled nonlinear systems with many degrees of
freedom (two weakly coupled low-amplitude and macroscopically wide
breathers) can be (approximately) mapped onto the dynamics of
exactly solvable nonlinear dimer. The (macroscopically) large
number of degrees of freedom in the considered coupled systems
also allows one to consider their relative phase $\Phi$ as an
observable quantity, like in the case of weakly coupled
Bose-Einstein condensates \cite{albiez}.

It is worth to mention that the form and corresponding frequency
of a breather in an $isolated$ $chain$ can be obtained in our
model only in the self-trapping breather regime, in which one can
consider the limit of $C\rightarrow 0$. Indeed, according to Eq.
(46), in this limit one has $\Theta=0$ ($z=1$) in Eqs. (19) and
(20), and the breather frequency is equal to
\begin{equation}
\omega(\Psi_{max},k) =\omega_{m}+(\frac{9}{8}\beta
-2\alpha^2)\Psi_{max}^2 +\frac{3}{8}\beta \Psi_{max}^2
-\frac{1}{4}k^2 = 2 +(\frac{3}{2}\beta -2\alpha)\Psi_{max}^2
-\frac{1}{4}k^2.
\end{equation}

This expression for the breather frequency is fully consistent
with the known expressions for (stationary or slowly-moving)
breather in a single $\alpha$-$\beta$-FPU chain in the
small-amplitude limit, see, e.g., Refs. \cite{koskov,kos3}. It is
important to emphasize that to get this expression for breather
frequency $\omega$, one has explicitly take into account the
linear increase in time ($winding$ $up$)  of the relative phase
$-\Phi/2$ in Eq. (19), given by Eq. (46) in the self-trapping
regime. Similar winding up of the relative phase of two coupled
macroscopic Bose-Einstein condensates in the nonlinear
self-trapping regime has been recently measured  directly  in a
single bosonic Josephson junction
 by means of interference \cite{albiez}. This finding gives us an additional argument in favor of the
 similarity between macroscopic tunneling quantum dynamics and phase-coherent dynamics of weakly coupled
 breathers.

\section{Numerical simulations and comparison with analytical predictions}

To simulate the dynamics of coupled nonlinear lattice excitations,
we numerically integrate equations (2) for two identical chains,
$l^{(1)}=l^{(2)}=1$, $\alpha^{(1)}=\alpha^{(2)}\equiv\alpha$ and
$\beta^{(1)}=\beta^{(2)}\equiv\beta$, with absorbing edge
conditions. We use the latter conditions in order to get rid of
weak radiation, caused by the wandering breather (because such
radiation will stay in the system forever in the case of periodic
boundary conditions). For the convenience of the numerical
simulations, the Fermi-Pasta-Ulam Hamiltonian (1) for two coupled
chains with $N$ particles can be written in the following
equivalent dimensionless form:
\begin{equation}
H=\sum_{i=1}^{2}\left[\sum_{n=1}^N\frac{1}{2}\dot{u}_{n}^{(i)2}
+\sum_{n=1}^{N-1} V(u_{n+1}^{(i)}-u_{n}^{(i)})
+\sum_{n=1}^N U(u_{n}^{(i)}-u_{n}^{(3-i)})\right], \label{f1}
\end{equation}
where the potentials $V(x)$ and $U(x)$ describe, respectively, the
intra- and inter-chain interactions. We normalize the
dimensionless potential $V(x)$  with the conditions
$V(0)=V'(0)=0$, $V''(0)=1$, while the potential $U(x)$ we
normalize  with the conditions $U(0)=U'(0)=0$, $U''(0)=C>0$.
According to Eq. (1), we take the following form for the intra-
and inter-chain potentials:
\begin{eqnarray}
 \label{f2}
V(x)&=&\frac12 x^2+\frac13\alpha x^3+\frac14\beta x^4, \\
U(x)&=&\frac12 C x^2, \label{f3}
\end{eqnarray}
with $C=0.1$ describing the strength of the weak inter-chain coupling.

Then the equations of motion (2) will take the form as
\begin{equation}
\ddot{u}_{n}^{(i)}=F(r_{n}^{(i)})-F(r_{n-1}^{(i)})+G(\delta_{n}^{(i)}),
\label{f6}
\end{equation}
where $r_{n}^{(i)}=u_{n+1}^{(i)}-u_{n}^{(i)}$,
$\delta_{n}^{(i)}=u_{n}^{(3-i)}-u_{n}^{(i)}$, $i=1,2$,
$F(r)=V'(r)=r+\alpha r^2+\beta r^3$,
$G(\delta)=U'(\delta)=C\delta$.

We search for the localized nonlinear lattice excitation (discrete
breather) as a solution of a set of nonlinear equations ${\cal
F}({\bf X})={\bf 0}$, where the vector ${\bf
X}=\{u_{n}^{(i)}(0),~\dot{u}_{n}^{(i)}(0)\}_{i=1, n=1}^{2,N}$
gives the initial values for Eqs. (\ref{f6}), while the vector
${\cal F}({\bf X})=\{u_{n}^{(i)}(t_p)-u_{n}^{(i)}(0) ,
~\dot{u}_{n}^{(i)}(t_p)-\dot{u}_{n}^{(i)}(0)\}_{i=1, n=1}^{2,N}$
gives the change of the vector ${\bf X}$ during the one period of
breather oscillations $t_p=2\pi/\omega$. To find the value of
${\cal F}({\bf X})$, one needs to integrate numerically Eqs.
(\ref{f6}) during the time interval $[0,t_p]$. The use of such
numerical method of finding of an exact breather solution is
explained in details in Ref. \cite{flach}. The main difficulty in
this numerical method is to find an appropriate initial vector
${\bf X}$ for the subsequent iterative solution of nonlinear
equations (\ref{f6}).

The energy of the breather is determined correspondingly as
\begin{equation}
E=\sum_{i=1}^{2}\left[\frac{1}{2t_p}\sum_{n=1}^N\int_{0}^{t_p}[\dot{u}_{n}^{(i)2}+V(r_{n}^{(i)})+V(r_{n-1}^{(i)})+\frac12 C\delta_{n}^{(i)2}]dt\right]\equiv\sum_{i=1}^{2} E_i . \label{f10}
\end{equation}

\subsection{Simulation of dynamics of discrete breathers in two weakly coupled chains}

Equations (53) for two coupled chains can be reduced to the
equations for one chain in the case of symmetric, $u_{n,1}\equiv
u_{n,2}\equiv u_n$, and antisymmetric, $u_{n,1}=-u_{n,2}\equiv
u_n$, motion in the chains. In the symmetric case the Hamiltonian
(50) has a usual form for a single FPU chain:
\begin{equation}
H=2\sum_{n=1}^N\{\frac12\dot{u}^2_n+V(u_{n+1}-u_n)\}. \label{f11}
\end{equation}

For the symmetric intra-chain potential $V(r)$, i.e. for
$\alpha=0$, equations of motion in a system of two chains can also
be reduced to the one-chain equations in the case of antisymmetric
(anti-pase) motion in the chains. In this case the reduced system
has a standard Hamiltonian of the FPU chain on an external
harmonic substrate:
\begin{equation}
H=2\sum_{n=1}^N\{\frac12\dot{u}^2_n+V(u_{n+1}-u_n)+2C u_n^2\}. \label{f12}
\end{equation}

Discrete breathers in such chain with $C\ge 0$ were studied in
details, see, e.g., Ref. \cite{flach}. In the system with
Hamiltonian (\ref{f12}), either with $C=0$ or $C>0$,  there are
two types of discrete breathers for
$\omega>\omega_{max}\equiv\sqrt{4+2C}$, see, e.g, Refs.
\cite{dolgov,siev,page}. The center of symmetry of the breather
coincides with a given lattice cite in one breather type, and it
coincides with a middle point of two neighboring cites in another
type. In both cases of the symmetric and antisymmetric breathers
in the coupled chains, the vibrational energy of the nonlinear
excitation, Eq. (\ref{f10}), is equally distributed between the
chains. Numerical analysis of Eqs. (\ref{f6}) reveals that besides
these (trivial) nonlinear eigenmodes there is an ``asymmetric''
mode, in which only one chain is involved in the vibration while
the vibration of the other chain ``adiabatically'' follows the
vibration of the main one. The characteristic form of the
displacement patterns in one- and two-chain antisymmetric
breathers is shown in Fig. 2, while the dependence of their
energies on breather frequency $\omega$ is shown in Fig. 3.

As we can see from the comparison of Figs. 2(c) and 2(d), the
magnitudes of the static displacement kinks in the center-of-mass
displacement differences  $\Delta
w_{0}^{(CM)}=w_{0}^{(CM)}(+\infty)-w_{0}^{(CM)}(-\infty)$ differ
almost twice for the two- and one-chain breathers in coupled
chains with cubic intra-chain anharmonic potential, in accordance
with the prediction of Eqs. (9) and (10) that the ratio of such
magnitudes is given mainly by the ratio of the total vibration
energies of the corresponding breathers. As follows from Fig.
3(c), the ratio of the (total vibration) energies of the two- and
one-chain breathers for $\omega=2.3$, $\alpha=-1$ and $\beta=1$ is
indeed close to two.

In the absence of the cubic anharmonic intra-chain potential
($\alpha=0$, $\beta=1$), the energy of the antisymmetric two-chain
breather monotonously increases from zero, for
$\omega=\omega_{max}=\sqrt{4+2C}\approx 2.05$, to the infinity for
$\omega\rightarrow \infty$, see Fig. 3(a). For low frequencies
$\omega_{max}<\omega< 2.066$, there is only one type of the
antisymmetric two-chain breathers, but at
$\omega=2.066\equiv\omega_{bif}$ an one-chain breather mode
detaches from the two-chain mode. The one-chain mode exists for
all frequencies  $\omega>\omega_{bif}$. As follows from Figs. 3(a)
and 3(b), there is a typical bifurcation of the nonlinear mode,
cf., e.g., Ref. \cite{akhmediev}.

We can get an analytical estimate for the characteristic frequency
 $\omega_{bif}$ with the use of generic evolution equations (27)
and (28). Indeed, for positive $\beta$, $\Phi=\pi$ and
$\omega_{m1}=\omega_{m2}$ one has an equation for the
steady-state, with $\dot{\Theta}=0$ and $\dot{\Phi}=0$, dynamical
state in the coupled chains,
\begin{equation}
 \sqrt{1-z^2}=\frac{C}{3\beta\Psi_{max}^2},
 \end{equation}
which has a solution for real $z$ only for $3\beta\Psi_{max}^2\geq
C$. Using this condition and Eq. (36) for breather frequency, in
the case of $\alpha=0$ and $k=0$, we get the following expression
for $\omega_{bif}$ at the appearance of the new solution (for
$C=0.1$):
\begin{equation}
 \omega_{bif}=\sqrt{4+\frac{5}{2}C}=\sqrt{4.25}=2.062,
 \end{equation}
which coincides with good accuracy with the numerically revealed
value $\omega_{bif}=2.066$. As follows from Eq. (57), at the
bifurcation point $3\beta\Psi_{max}^2 =C$ one has $z=0$. This
corresponds to the equal populations of the coupled chains, which
in turn results in equal energies of the two- and one-chain
breather solutions at the bifurcation, see Fig. 3 (a).

Two types of the coupled breathers, the one- and two-chain
breathers, can exist also in the presence of cubic term in
anharmonic intra-chain potential  $V(x)$, Eq. (51). The effect of
the cubic term in anharmonic intra-chain potential on breather
dynamics in a single FPU chain was studied in Refs.
\cite{koskov,bickham,huang,kiselev,kastner}. For instance, it was
shown in Ref. \cite{koskov} that low-amplitude breathers can exist
in a single FPU chain only in the case of relatively weak cubic
anharmonic term in the intra-chain potential $V(x)$, namely when
$\alpha<\sqrt{3\beta/4}$. (For the considered case of $\beta=1$,
this requirement reduces to $\alpha<\sqrt{3/4}=0.866$).  In fact
this requirement corresponds to the requirements
$\omega>\omega_{m}$ in Eq. (49) and $\lambda_1>0$ in Eq. (23) for
breather frequency and inverse localization length in a single FPU
chain (for $\Theta=0$ and $k=0$).

From our analysis we can conclude that  bifurcation of breather
modes in two weakly coupled nonlinear chains with cubic
intra-chain anharmonic potential occurs only for
$\alpha<\sqrt{3\beta/8}$ (or $\alpha<0.61$ in the considered
case). For larger cubic intra-chain anharmonic potential
$\alpha>\sqrt{3\beta/8}$, there are no low-amplitude breathers in
the coupled chains but there are finite-amplitude breathers with a
threshold in energy which does not go to zero for $\omega
\rightarrow \omega_{max}+0$, see Fig. 3(c) and cf. with Fig. 4 in
Ref. \cite{kastner}. Indeed, from Eq. (23) follows that for the
wandering breather, when
$\langle\cos^{2}\Theta\textsc{}\rangle=\langle\sin^{2}\Theta\textsc{}\rangle=0.5$,
the inverse localization lengths $\lambda_{1,2}$ are real positive
only for $\alpha<\sqrt{3\beta/8}$, when the above picture of the
breather periodic translation (wandering)  between the coupled
chains takes place. On the other hand,  for
$\alpha>\sqrt{3\beta/8}$ (or $\alpha>0.61$ in the considered
case), one has to take into account the contribution to breather
frequency $\omega$ and inverse square localization lengths
$\lambda_{1,2}^2$ of  the finite-amplitude breather of the terms
$\propto\Psi_{max}^{4}$ (with positive coefficients). These terms
will result in a finite (threshold) energy of the breather with
minimal possible frequency $\omega_{min}>\omega_{m}$. According to
Eqs. (14), the existence of such breather threshold amplitude and
energy in the coupled chains with $\alpha>\sqrt{3\beta/8}$ is
related with the static strain in both chains ($\partial
w_{0}^{(CM)}/\partial x =-\alpha\langle \mid\Psi_{1}\mid^2
+\mid\Psi_{2}\mid^2\rangle$), which accompanies breathers in such
coupled chains and which is localized around the central region of
the breather, see Eqs. (9) and (10) and Figs. 2(c) and 2(d).

In the coupled chains with $\alpha>\sqrt{3\beta/8}$, there is no
bifurcation of the two-chain breather mode: there are two
separated breather modes, the one- and two-chains ones, for all
the frequencies  $\omega> \omega_{min}$, the energy of the
two-chain breather mode is always larger than the energy of the
one-chain one, and there are thresholds in energies for both types
of breathers, see Fig. 3c. In the following we will restrict
ourselves only to the case of weakly coupled $\beta$-FPU chains
with zero cubic anharmonic intra-chain potential and assume
everywhere below that $\alpha=0$ and $\beta=1$.

For the large-amplitude (high-frequency) nonlinear oscillations in
two  weakly coupled $\beta$-FPU chains, there are many coupled
(bound) antisymmetric breather modes whose strongly localized
displacement patterns are shifted one with respect to another
along the chains, see Figs. 4(a) - 4(j). In these figures the
total breather energy $E$, Eq. (54), location of the breather
center in the corresponding chain, and the "binding  energy", i.e.
the gain in energy due to breather coupling, $\Delta E=
E_{1}^{(1ch)}+E_{2}^{(1ch)} -E$, are indicated for breathers with
(high) frequency $\omega=5$. Here $E_{1,2}^{(1ch)}$ indicate the
energy of the corresponding one-chain breather located in the
corresponding chain, see Figs. 4(a) and 4(b) for the two main
types of the one-chain breathers: with the center of symmetry
placed in the middle of two neighboring sites or in a particular
site. As it is seen from Fig. 4, the binding energy depends on the
type of the bound breathers and apparently decreases with the
increase of the inter-breather distance (along the chains).
Besides such one-frequency coupled breather modes, there are also
two-frequency nonlinear modes which bind together large-amplitude
breather modes in two chains with different but $commensurate$
$frequencies$ $\omega_1$ and $\omega_2 >\omega_1$. The dependence
of the energy of several types of such commensurate-frequency
bound breather modes on their maximal frequency $\omega_2$ is
shown in Fig. 5.

We also study the interaction of breathers in two weakly coupled
$\beta$-FPU chains. For this purpose, we numerically integrate
equations of motion (53) with the initial conditions which
describe a breather with frequency $\omega_1$ on the first chain
and a breather with frequency $\omega_2$ on the second chain. In
Fig. 6  we show the change in time of the spectral density
$p(\omega)$ of the emerged localized oscillation mode in the
coupled neighboring chains with two breathers with initial
frequencies $\omega_1=2.6$ and $\omega_2=3$. As one can see in
this figure, a two-frequency breather is generated in the system
with the ratio of frequencies $\omega_1:\omega_2=6:7$, which
subsisted for 1200 units of dimensionless time (when time is
measured in units in which $\omega_{m}=\sqrt{4+C}\approx 2$).
Later the breather with higher frequency (and energy) starts to
adsorb the energy of the breather with smaller frequency (and
energy), which decreases the frequency of the latter. In result,
the one-chain breather with frequency $\omega=3.160$ is formed.
Hence the breather with higher energy completely adsorbs the
breather with smaller energy. This reminds the $merging$ of
low-amplitude (small-energy) breathers in a single 1D $\beta$-FPU
chain, which results in a single localized object (discrete
breather) containing almost all of the initial total vibration
energy of the lattice \cite{kos3}. If we initially prepare two
breathers with more close initial frequencies $\omega_1=2.7$ and
$\omega_2=3$, a two-chain one-frequency breather with
antisymmetric displacement pattern is formed (not shown). In this
case the interaction and energy exchange of the two breathers with
initially different frequencies result in frequency equalization.
Therefore our simulations show that despite the existence of the
coupled two-frequency breathers, the one-frequency coupled
breather modes are in general more stable.

\subsection{ Simulation of wandering breather in two weakly coupled chains}

Here we will study the dynamics of two weakly coupled chains under
excitation of a breather only in one chain while keeping the other
chain initially at rest. This amounts in numerical integration of
Eqs. (53) for chain 1 under the initial conditions $u_{n}^{(2)}=0$
and  $\dot{u}_{n}^{(2)}=0$. In result of inter-chain interaction,
the initially located in one chain excitation can start to
periodically translate (wander) between the coupled chains. Such
inter-chain energy exchange can be studied quantitatively by
measuring the time dependence of the energy $E_i$ in each chain,
see Eq. (54).

In Fig. 7 we plot the energy of each chain $E_i$ as a function of
time when the immovable breather (with $V=0$) is excited in chain
1 with a given frequency $\omega$. For relatively small breather
amplitude and corresponding dimensionless breather frequency,
$\omega=2.030>\omega_{m}=\sqrt{4+C}=2.025$, an almost complete
inter-chain energy exchange occurs when the total breather energy
$E=E_1 +E_2$, Eq. (54), is periodically located in one of the
coupled chains, see Fig. 7(a). From this figure we can measure the
period of wandering $T_{w}$, which is close to 126. Taking into
account that the frequency of the complete inter-chain energy
exchange for small-amplitude breather is equal to $C/2=0.05$, see
Eqs. (21) and (45), the wandering period $T_{w}$ is equal to
$4\pi/C=125.66$, which coincides with very high accuracy with the
period, which can be found from Fig. 7(a). With the increase of
breather amplitude and frequency, the period of the inter-chain
energy exchange increases, in accordance with Eq. (45), but the
exchange remains complete up to the critical breather frequency
(and corresponding amplitude), which according to Figs. 7(b) and
7(c) is very close to  $\omega=2.0980$. This critical breather
frequency should be compared with the value of frequency of the
immovable breather at the separatrix, $\omega_{sep}
=\sqrt{4+4C}=2.0976$ for $C=0.1$, $\alpha=0$ and $k=0$, see Eq.
(48), which also shows very good agreement with the prediction of
the approximate physical-pendulum equation (39) (or (40)). For the
higher breather frequencies $\omega> 2.098$ (and corresponding
breather amplitudes), the inter-chain energy exchange is no more
complete and corresponds to the exchange of relatively small
fraction of the total breather energy $E$, see Figs. 7(c) and
7(d). For $\omega=2.1106$, the main part of the breather energy
remains all the time in chain 1 were the breather was initially
excited, which corresponds to nonlinear self-trapping of the
breather, see Fig. 7(d). The transition from the oscillatory
energy exchange, Fig. 7(b), to the self-trapped mode, Fig. 7(c),
is very sharp: the change of the breather frequency by the
crossing of the separatrix occurs in the fifth digit only. One can
also see a big difference in energy exchange rate in the wandering
breather mode, Fig. 7(a), and in the separatrix mode, Figs. 7(b)
and 7(c): with the change of the breather frequency from
$\omega=2.0300$ to $\omega=2.098$, the rate of energy exchange
decreases more than twenty times! This observed feature of the
coupled nonlinear systems is also in accordance with the
prediction of the physical-pendulum equation (39) (or (40)) for
the separatrix mode with the infinite period of the inter-chain
energy exchange, Eq. (47).

In Fig. 8 we show the Fourier power spectrum of the wandering
breather. As it was explained above in connection with complex
wave field of the wandering breather, Eqs. (19) and (20), the
temporal Fourier spectrum of wandering breather consists of the
main peak at breather fundamental frequency $\omega$ and its
(upper and lower) frequency satellites at $\omega\pm
n\omega_{beat}$, $n=1,2,...$, where $\omega_{beat}\approx
(C-9\beta^2\Psi_{max}^4/(16C))/2$ is the rate of complete
inter-chain energy exchange in the wandering breather mode, see
Eq. (45). The value of $\omega_{beat}\approx 0.05$ for
low-amplitude wandering breather in the chains with the coupling
constant $C=0.1$ is in a good agreement with the inset in Fig. 8.
Therefore we can quantitatively reproduce both the numerically
observed period, Fig. 7(a), and wandering (beating) frequency,
Fig. 8, of the low-amplitude and macroscopically wide breather.

The nonlinearity of the physical-pendulum Eq. (39) (or (40))
results in all higher harmonics (second, third etc.) of the
beating frequency $\omega_{beat}$, and respectively in the
appearance of corresponding doublets of breather frequency
satellites: the satellites of the first and second harmonics of
the beating frequency are clearly seen in the inset in Fig. 8. As
we can also see in the inset in Fig. 8, the lower-side satellites
enter the continuum of low-amplitude phonons,
$\omega-n\omega_{beat}<\omega_{m}=\sqrt{4+C}=2.025$, $n=1,2,...$.
The latter means that the wandering breather (weakly)  emits
low-amplitude lattice phonons, which results in (adiabatically
slow) lowering of its energy. This in turn means that the
wandering breather is not an exact nonlinear eigenmode of two
weakly coupled anharmonic chains, in contrast to the two-chain
symmetric and antisymmetric breathers.

In Fig. 8 one can also see the third harmonic of the main breather
frequency $\omega$, which can be modelled with the use of the
full-harmonics Eqs. (12) or (53), instead of Eqs. (14) written in
the resonant approximation. For the $\beta$-FPU chain (with zero
cubic anharmonic intra-chain potential), there are only odd
harmonics of the fundamental breather frequency. In  a single
chain, the higher harmonics of the fundamental breather frequency
were first analytically predicted and described in Ref.
\cite{koskov}, and were numerically observed, up to the ninth
harmonic, in Ref. \cite{kos4}. In Fig. 9 we show the power
spectrum of the one-chain breather with frequency $\omega=4$ in
the coupled chains with $\alpha=0, \beta=1, C=0.1$. This spectrum
clearly shows the existence of higher odd harmonics, up to the
seventh one, of the fundamental frequency. Since the breather with
such frequency is in the self-trapped mode  in the considered
coupled system ($\omega=4>\omega_{sep}=2.1$), its dynamics and
power spectrum are similar to the ones of the breather in a single
chain, cf. Refs. \cite{kos3,kos4}. Importantly, the Fourier
spectrum in Fig. 9 corresponds to almost ``exact'' breather
solution found in the $\beta$-FPU chain with zero vibrational
background, while the breather, studied in Refs. \cite{kos3,kos4},
was self-created on a non-zero vibrational background in result of
modulational instability of short-wavelength modes in the
$\beta$-FPU chain.

In view of the present work, we can relate the observed in Ref.
\cite{kos4} doublets of satellites of the fundamental breather
frequency (and its odd harmonics) in a $single$ $chain$ with the
beating between two breather states with different center
locations (and symmetry), see Figs. 4(a) and 4(b). According to
these figures, the breather with the inter-site center location,
Fig. 4(a), has higher frequency than the breather with the same
energy with the on-site center location, Fig. 4(b). According to
generic Eq. (28), the frequency difference under the proper
initial conditions can cause breather wandering (beating) between
two nearest lattice sites. The beating in turn can induce the
steady-state translation of the breather along the chain with
(slow) velocity $V\ll 1$, proportional to the (small) beating
frequency (as it was qualitatively explained in Ref. \cite{kos4}).
In Section III C we will discuss similar lateral translation of 1D
breather in a system of parallel weakly coupled nonlinear chains
with transverse group velocity, proportional to the beating
frequency in the case of two coupled chains. In contrast to the
low-amplitude 1D breather, laterally moving and losing its energy
due to phonon emission in a system of nonlinear chains, see Figs.
14(a) and 14(c) below, the high-frequency slowly-moving breather
in a single chain does not emit phonons in the chain because the
fundamental frequency satellites (caused by the beating) are all
out of the phonon band, see Fig. 3 in Ref. \cite{kos4}.

Similar picture of the complete inter-chain energy exchange takes
place also for wandering breather, $slowly$ $moving$ along the
chains. To model this effect, we numerically integrate Eqs. (53)
with the initial conditions which correspond to the excitation of
the moving in chain 1 breather solution, while keeping chain 2 at
rest, with the following ansatz for lattice displacements:
\begin{equation}
u_{n}^{(1)}(t)=(-1)^n \Psi_{max}\cos(kn-\omega t)/\cosh[\lambda
(n-Vt)],~~~ u_{n}^{(2)}(0)=0,~~~ \dot{u}_{n}^{(2)}(0)=0,
\label{f13}
\end{equation}
where  $\Psi_{max}$, $\omega=\omega(\Psi_{max},k)$ and $V\approx
-k/\omega\ll 1$ are the amplitude, frequency and slow velocity of
the breather, cf. Eq. (19).

In Figs. 10 and 11 we show the energy, versus time and site, of
slowly-moving breather,  with velocity $V=0.1$, wandering between
two weakly coupled chains. In Fig. 10 we show the energy of the
wandering small-amplitude breather below the separatrix: with
$\Psi_{max}=0.1$, Figs. 10(a) and 10(b), and $\Psi_{max}=0.2$,
Figs. 10(c) and 10(d). Both figures show the complete inter-chain
energy exchange. From Figs. 10(a) and 10(b) one can find that the
period of the complete inter-chain energy exchange (wandering)
$T_{w}\approx 128$ for $\Psi_{max}=0.1$ is very close to the one
$T_{w}\approx 126$ in Fig. 7: both figures describe almost
harmonic wandering of a small-amplitude breather, with wandering
frequency $C/2=0.05$. Below the separatrix, the wandering period
$T_{w}$ increases with breather amplitude: one has $T_{w}\approx
179$ for $\Psi_{max}=0.2$ in Figs. 10(c) and 10(d). Figure 11
shows the dynamics of slowly-moving breather initially excited in
chain 1 with amplitude $\Psi_{max}=0.26$, which is very close to
the one at the separatrix:
$\Psi_{max}^{(sep)}=\sqrt{2C/\beta}=0.2582$. Separatrix-like
dynamics, similar to the one shown in Fig. 7(b), is well
established for the later delay time $t$$\geq$4000, when almost
total energy of the moving breather periodically translates
(wanders) between the chains. In Fig. 12 we show the dynamics of
slowly-moving breather, initially excited in chain 1 with velocity
$V=0.1$ and amplitude $\Psi_{max}=0.3$, which is beyond the
separatrix. Here the inter-chain energy exchange is no more
complete, like in the case of immovable breather beyond the
separatrix, cf. Fig. 7(d), but the period of such exchange,
$T_{x}\approx 125$, is much shorter than that close to the
separatrix, cf. Fig. 11.

Slowly-moving $wandering$ $Bose$-$Einstein$ $condensate$  can be
realized in two coupled 1D atom waveguides (see, e.g., Refs.
\cite{kinosh,wang}), for the BEC of weakly-interacting atoms with
non-negligible inter-waveguide tunneling coupling and with total
initial population imbalance $\mid$$z(0)$$\mid$=1.

Therefore the excitation of the low-amplitude (low-frequency)
breather, either immovable or slowly-moving,  in one chain always
results in its periodic transverse translation (wandering) between
the two weakly coupled chains. Hence a natural question arises
what happens if one deals with a system of $M>2$ parallel coupled
anharmonic chains. Whether the wandering of the breather,
initially excited at the edge (outermost) chain, across all the
chains is
 possible? To study this problem, we start with a system of $M>2$
parallel weakly coupled anharmonic chains.

\subsection{Simulation of breathers in a system of $M>2$ parallel weakly coupled anharmonic chains}

Quasi-1D system of $M$ parallel weakly coupled anharmonic chains,
with nearest-neighbor intra- and inter-chain interactions, is
described by the following Hamiltonian:
\begin{equation}
H=\sum_{m=1}^M\sum_{n=1}^N\frac12\dot{u}_{m,n}^2
+\sum_{m=1}^M\sum_{n=1}^{N-1} V(u_{m,n+1}-u_{m,n})
+\sum_{m=1}^{M-1}\sum_{n=1}^N U(u_{m+1,n}-u_{m,n}), \label{f1}
\end{equation}
where $V(x)$ and $U(x)$ are given by Eqs. (51) and (52), and
$n=1,...,N$ and $m=1,...,M$ numerate, respectively, sites along
the chains and the chains.

Hamiltonian (60) generates corresponding equations of motion,
\begin{equation}
\ddot{u}_{m,n}=-\frac{\partial H}{\partial u_{m,n}},
\label{f14}
\end{equation}
which in the linear approximation have the form:
\begin{equation}
\ddot{u}_{m,n}=u_{m,n+1}-2u_{m,n}+u_{m,n-1}+C(u_{m+1,n}-2u_{m,n}+u_{m-1,n}).
\label{f14}
\end{equation}
 Plane linear waves (phonons) in such system, with
\begin{equation}
u_{m,n}=u\exp[iq_1 n +iq_2 m -i\omega t],
\label{f14}
\end{equation}
have the dispersion:
\begin{equation}
\omega(q_1,q_2)=\sqrt{2[1-\cos q_1+ C(1-\cos q_2)]}, \label{f15}
\end{equation}
where both the intra-chain lattice period and inter-chain spacing
are taken equal to unit, and therefore $0\le (q_1,q_2)\le \pi$.
Minimal phonon frequency in this translationally-invariant system
is zero, $\omega(0,0)=0$, while cut off phonon frequency is
$\omega(\pi,\pi)=2\sqrt{1+C}$ (which is equal to
$\omega(\pi,\pi)=2.0976\approx 2.1$ for  $C=0.1$).

Since we will be interested in the short-wavelength excitations,
with $q_1\approx\pi$, the corresponding phonon frequency for $C\ll
1$,
\begin{equation}
\omega(q_1\approx\pi,q_2)=2+C\sin^2 (\frac{1}{2}q_2), \label{f16}
\end{equation}
determines the phonon group velocity across the chains:
\begin{equation}
V_{\perp}(q_1\approx\pi,q_2)=\frac{\partial\omega(q_1\approx\pi,q_2)}{\partial q_2}=\frac{1}{2}C\sin(q_2). \label{f17}
\end{equation}

Now we turn to the nonlinear dynamics of $M$ weakly coupled
parallel nonlinear chains with Hamiltonian (60). We will integrate
Eqs. (61) with the initial condition which describes exact
discrete breather in the $m$-th chain ($1\le m\le M$) under the
condition of immovability of the rest of the chains, to study the
time dependence (for $t>0$) of the vibration energy in the chains:
\begin{equation}
E_m=\frac12\sum_{n=1}^N(\dot{u}_{m,n}^2+V_{m,n}+V_{m,n-1}+U_{m,n}+U_{m-1,n})\equiv\sum_{n=1}^N
E_{n,m},
\end{equation}
where $V_{m,n}=V(u_{m,n+1}-u_{m,n})$,
$U_{m,n}=U(u_{m+1,n}-u_{m,n})$, and breather energy
$E=\sum_{m=1}^M E_{m}=\sum_{n=1}^N\sum_{m=1}^M E_{n,m}$.

By measuring the time dependence of $E_{n,m}$, we can define the
breather center location $(n_c,m_c)$,
\begin{equation}
n_c=\sum_{m=1}^M\sum_{n=1}^N
np_{n,m},~~m_c=\sum_{m=1}^M\sum_{n=1}^N mp_{n,m},
\end{equation}
where $p_{n,m}=E_{n,m}/E$ describes the distribution of normalized
breather energy in 2D lattice. With these breather quantities we
can measure the longitudinal $D_x$ (along the chain axis) and
transverse $D_y$ breather diameters:
\begin{equation}
D_x=2\left[\sum_{m=1}^M\sum_{n=1}^N
(n-n_c)^2p_{n,m}\right]^{1/2},~~~
D_y=2\left[\sum_{m=1}^M\sum_{n=1}^N (m-m_c)^2p_{n,m}\right]^{1/2}.
\end{equation}

 In Figs. 13
(a,b,c,d) we show the time dependence of energies of the first,
$i=1$, and the last, $i=M$, coupled chains for $M=2,3,4,5$, for
the time interval just after breather excitation in the first
chain, left panel, and for the later time, right panel. In the
case of $M=2$, there is a periodic (harmonic) complete energy
exchange between the first and the second chains, see Fig. 13(a).
In the case of $M=3$, there is periodic (non-harmonic) recurrence
of the complete energy accumulation in the first chain, and the
period of such recovery is twice larger (the recurrence rate is
twice smaller) than that in the case of $M=2$. For $M=3$, the time
dependence of the first chain energy recurrence can be roughly
approximated as $\cos^{4}(Ct/8)$ (instead of $\cos^{2}\Theta
=\cos^{2}(Ct/4)$ for $M=2$, Eqs. (19) and (45)), similar to the
time dependence of the population of the initially excited state
in a three-state (spin 1) atomic system, see, e.g., Ref.
\cite{mewes}. Here $C/4$ plays role of the Rabi frequency, which
is twice smaller than the rate of the complete energy exchange
(energy recurrence) in the case of $M=2$, when it is equal to
$C/2$, see Eq. (45). In the case of $M=4$ and $M=5$, the
recurrence of energy of the first and the last chains becomes
quasi-periodic, see Figs. 13 (c) and 13 (d), but still the
(approximate) period $T_{M}$ of such recurrence scales with the
number of chains $M$ as $T_{M}\propto(M-1)$. For $M\geq 6$,
$C=0.1$, the initially localized (in chain 1) excitation spreads
its energy in the whole system of weakly coupled chains.

The dependence of the recurrence period $T_{M}$ on $M$ we can
relate with the transverse group velocity, Eq. (66), of the
high-frequency phonons (with $q_1\approx\pi$) in the system of
weakly coupled anharmonic chains. The transverse wavevector $q_2$
in Eq. (66) is equivalent to the relative phase of the neighboring
chains. In the regime of almost-harmonic energy transfer between
the neighboring chains, the relative phase is always close to
$\pi/2$, see Fig. 1. It means that the transverse wavevector $q_2$
in the expression (66) for the transverse group velocity of the
wandering breather should also be  (approximately) equal to
$\pi/2$. In the case of 2D system of coupled oscillator chains,
$q_2=\pi/2$ corresponds to the case when, say,  only the odd
chains are excited at a given moment while their nearest
neighbors, even chains, are at rest. This gives
$V_{\perp}\approx\frac{1}{2}C$ for the transverse group velocity.
Importantly, this characteristic group velocity does not depend on
the number of the coupled chains $M$. Therefore we can estimate
the period of the first chain energy recovery as
$T_{M}=2A(M-1)/V_{\perp}\approx 4A(M-1)/C$ with some dimensionless
factor $A$, which is consistent with our numerical observation
(with $A\approx 3$).

The same transverse phonon group velocity one can estimate from
Fig. 14 as the speed of breather spreading across the chains.
Figure 14 shows the time dependence of energy distribution among
the chains when the breather is initially excited in the edge
chain, Figs. 14(a) and 14(b), or in the central chain, Figs. 14(c)
and 14(d), in the system of $M=50$ coupled chains with $N=50$. As
follows from Figures 14(a) and 14(c),  the initial breather energy
has spread for 20 chains for approximately 500 time units. This
gives us quantitative estimate of 0.04 for the transverse group
velocity, which is rather close to our analytical estimate with
the use of Eq. (66) for $q_2\approx\pi/2$:
$V_{\perp}\approx\frac{1}{2}C=0.05$. Figure 14 also shows that the
appearance of localized breather in a system of coupled chains,
with $M\gg 2$, has a threshold in breather frequency,
$\omega_{thresh}=2.15$ in Fig. 14(b) and $\omega_{thresh}=2.17$ in
Fig. 14(d), similar to the case of two coupled chains. The
threshold  breather frequency, which corresponds to the appearance
of localized breather in a system of coupled chains, we should
compare with breather frequency at the separatrix in a system of
two coupled chains, $\omega_{sep}\approx 2.1$, see Eq. (48) for
$C=0.1$, $\alpha=0$, $k=0$ and Figs. 7(b), 7(c) and 11. Since all
the above frequencies are close, we can get the estimate for the
threshold breather amplitude for its localization in 2D system of
weakly coupled chains: $\Psi_{max}^{thresh}\sim \sqrt{C/\beta}$.
For $\Psi_{max}<\Psi_{max}^{thresh}$, the 1D breather, which was
initially excited in one chain, will start to translate laterally
to the neighboring chains, will spread its energy among them and
lose it due to phonon emission, and finally will decay into
small-amplitude phonons due to lowering of its frequency up to the
cut off phonon frequency $\omega_{max}=2\sqrt{1+C}$, see Figs.
14(a) and 14(c). Such evolution of low-amplitude 1D breathers can
also be related with the above-mentioned conclusion that wandering
breather is not an exact solution of the nonlinear system even in
the case of two coupled anharmonic chains. In contrast to this
behavior of low-amplitude breathers, 1D breather with the
amplitude  $\Psi_{max}>\Psi_{max}^{thresh}$ is self-trapped and
remains localized mainly in the chain of its initial excitation,
see Figs. 14(b) and 14(d). [In Fig. 14 (d) one can also see a
partial (incomplete) energy exchange between the central chain and
its nearest neighboring chains, similar to the incomplete energy
exchange in two coupled chains beyond the separatrix, cf. Figs. 7
(d) and 12]. This phenomenon resembles the so-called
$delocalizing$ $transition$ in 2D system, when the wave field
abruptly changes its character from spatially localized to the
extended one, cf. similar delocalizing transition for polarons in
2D and 3D lattices \cite{kalosak} and for Bose-Einstein condensate
in 2D optical lattice \cite{kalosak2}. In our case, the
delocalizing transition occurs by the decrease of the initial
breather amplitude $\Psi_{max}$ (or frequency $\omega$) from the
value $\Psi_{max}>\Psi_{max}^{thresh}\sim \sqrt{C/\beta}$ (or
$\omega>\omega_{thresh}$) to the value
$\Psi_{max}<\Psi_{max}^{thresh}$ (or $\omega<\omega_{thresh}$).
Such transition is related with finite energy threshold for the
creation of solitons and breathers in 2D and 3D systems, see Refs.
\cite{kingsep,flach2}, and is absent in 1D ($\beta$-FPU or
discrete nonlinear Schr\"odinger equation \cite{kalosak,kalosak2})
systems. Indeed, the threshold breather amplitude
$\Psi_{max}^{thresh}$ in strongly anisotropic quasi-1D system
vanishes in the limit $C\rightarrow 0$ of a single 1D chain since
$\Psi_{max}^{thresh}\sim \sqrt{C/\beta}$. Similar threshold
breather amplitude $\Psi_{max}^{thresh}\sim \sqrt{C/\beta}$ for
breather delocalization should also appear in 3D array of parallel
weakly coupled nonlinear chains, with coupling constant $C\ll 1$,
which are described by the FPU Hamiltonian, similar to the one
given by Eq. (1).

To get more deep insight into the delocalizing transition of
discrete breather in 2D system of weakly coupled chains, we
confirm numerically that discrete breather in 2D system of weakly
coupled chains is stable only in the frequency range
$[\omega_b,\infty)$, where $\omega_b=2.12$ for $C=0.1$ is higher
than the cut off phonon frequency
$\omega(\pi,\pi)=2\sqrt{1+C}\approx 2.1$. This means that there is
frequency gap $[\omega(\pi,\pi),\omega_b]$ above the cut off
phonon frequency in which there are no solutions for localized
breathers in 2D system: breathers with frequencies in the gap
spread across all the 2D lattice.   The frequency $\omega_b$ is
close to the aforementioned and indicated in Fig. 14 threshold
frequencies $\omega_{thresh}$, but the frequency $\omega_b$ does
not depend on the way of the breather excitation in 2D lattice, in
contrast to the threshold frequencies. In Fig. 15 we plot the
breather energy distributions along and across the chains for the
one-chain, Fig. 15(a), and antisymmetric (anti-phase) two-chain,
Fig. 15(b), breathers in 2D lattice with $\omega=\omega_b=2.12$,
which show that transverse localization length of the discrete
breather has the order of the inter-chain spacing  for
$\omega=\omega_b$ (and further decreases for $\omega > \omega_b$,
see Fig. 16 (b) below).

In addition to the one- and two-chain breathers, there are also
three-chain, four-chain etc. in- and anti-phase breathers in a
multi-chain system. In Fig. 16 we show the vibration breather
energy and the measured (with the use of Eqs. (67)-(69))
longitudinal and transverse breather diameters for the one-, two-,
three- and four-chain anti-phase breathers. As one can see from
this figure, breather energy grows as $\omega^4$ in the
high-energy limit. This dependence can easily be understood from
Eqs. (16) and (22) for breather energy $E_b$ and frequency
$\omega$: one has $E_b\propto\Psi_{max}^4$ and
$\omega\propto\Psi_{max}$ in the large-amplitude limit, see also
Ref. \cite{kos4}. Breather with threshold frequency $\omega
=\omega_b$ has minimal energy: $E_b=0.517$ and $E_b=0.740$ for the
one- and anti-phase two-chain breathers, respectively. From Fig.
16 (b) we can also see that, in accordance with Fig. 15 (a), the
transverse localization length of the one-chain breather at the
threshold frequency is very close to the inter-chain spacing and
further decreases with the increase of breather energy. [Both
longitudinal and transverse breather diameters saturate at the
amplitude-independent values in the high-energy limit, see, e.g.,
\cite{kos2}, when the transverse diameter of the $m$-chain
breather simply is $D_{y}\approx B(m+1)$ with $B\cong 1$, see Fig.
16(b).] This means that there are no laterally-localized
envelope-soliton vibrational excitations in our 2D system of
parallel coupled nonlinear chains, in contrast to the
laterally-localized envelope-soliton optical excitations in 2D
arrays of parallel coupled nonlinear optical waveguides
\cite{eisen}. Therefore the inter-chain breather translation and
wandering are possible only in a system of small number of coupled
chains ($M\leq 5$ for $C=0.1$), while in a multi-chain 2D system
($M\geq 6$ for $C=0.1$) the vibrational breathers can move only
along the chains. Such 1D motion of 1D breathers in 2D system of
multiple parallel weakly coupled oscillator chains resembles the
1D motion of 1D Bose gases in 2D array of multiple parallel weakly
coupled atom waveguides (a quantum Newton's cradle) \cite{kinosh}
(see also \cite{winkler}).

\section{Summary}

 In summary, we have found, both analytically and
 numerically, two qualitatively different regimes of energy
exchange between phase-coherent breathers (intrinsically localized
short-wave nonlinear excitations)
 in two weakly linked nonlinear chains. In the low-amplitude mode, the breather performs  periodic transverse translations (wandering)
 between the coupled chains. In the large-amplitude mode, the
 breather is self-trapped in one chain. These two breather modes are detached by
 a separatrix, at which the rate of the inter-chain energy exchange
 vanishes.
 These regimes have a profound analogy, and are
 described by a similar pair of equations, to the
 Josephson plasma  oscillations and nonlinear self-trapping, recently observed in a single
 bosonic Josephson junction  \cite{albiez}.
The predicted evolution of the relative phase of two weakly
coupled coherent breathers in both regimes is also
analogous to the evolution of relative quantum mechanical phase
between two weakly coupled macroscopic condensates,  which was
directly measured  in a single bosonic Josephson junction by means
of interference \cite{albiez}. On the basis of this profound
analogy, we predict a new tunneling regime  of two weakly linked
Bose-Einstein condensates  in which  their relative phase
oscillates around $\frac{\pi}{2}$ modulo $\pi$, which can be
observed by means of interference. The  similarity between the
classical phase-coherent excitation exchange and macroscopic
tunneling quantum dynamics found here can encourage new
experiments in both fields.

We also show that static displacements of the coupled chains with
nonlinear localized excitation, induced by cubic term in the
intra-chain anharmonic potential, scales approximately as the
total vibrational energy of the excitation, either one- or
two-chain one, and do not depend on the inter-chain coupling. This
observation is also valid for a narrow stripe of several parallel
coupled nonlinear chains. We also study two-chain breathers, which
can be considered as bound states of discrete breathers with
different symmetry and center locations in the coupled chains, and
bifurcation of the anti-phase two-chain breather into the
one-chain one.  Bound states of two breathers with different
commensurate frequencies are found in the two-chain system.
Merging of two breathers with different frequencies in one
breather in two coupled chains is observed. Wandering of
low-amplitude breather in a system of several, up to five, coupled
nonlinear chains is studied, and the dependence of the wandering
period on the number of chains is analytically estimated and
compared with numerical results.

Delocalizing transition of 1D breather in 2D system of a large
number of parallel coupled nonlinear chains is described, in which
the breather, initially excited in a given chain, abruptly spreads
its vibration energy in the whole 2D system upon decreasing
breather frequency or amplitude below the threshold one. The
threshold breather frequency is above the cut off phonon frequency
in 2D system, the threshold breather amplitude scales as square
root of the inter-chain coupling constant, and breather vibration
energy is localized mainly in one chain at the delocalization
threshold. Similar delocalizing transition for 1D breather should
also occur in 3D array of parallel coupled nonlinear chains.
Delocalizing transition of discrete vibrational breather in 2D and
3D systems of coupled nonlinear chains has an analogy with
delocalizing transition for Bose-Einstein condensates in 2D and 3D
optical lattices.

\section*{Acknowledgements}

Yu. A. K. is grateful to S. Aubry, E. Bogomolny, S. Flach, V.
Fleurov, A. S. Kovalev  and G. Shlyapnikov for useful discussions.
This work was supported by the Presidium of the Russian Academy of
Sciences and Russian Foundation for Basic Research (Grants
04/BGTCh-07 and 05-03-32241).

\newpage

\begin{figure}[p]
\begin{center}
\includegraphics[width=7cm]{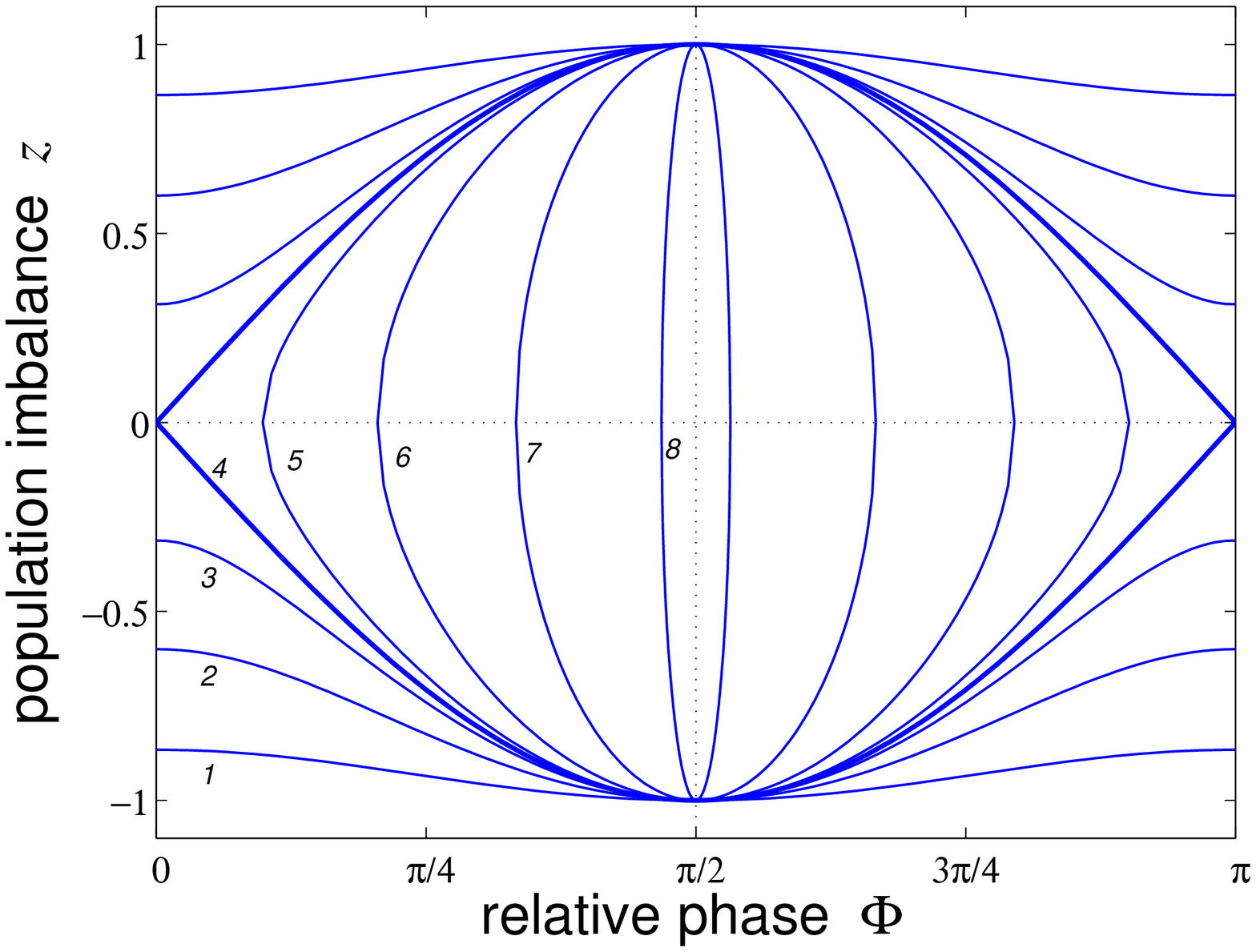}
\end{center}
\caption{\label{Fig1}\protect Phase portrait of wandering breather
in two weakly coupled nonlinear chains or two weakly linked BECs
in a symmetric double-well potential with the initial conditions
$\mid$$z(0)$$\mid$=$1$ and $\Phi(0)$=$\pi/2$ modulo $\pi$, which
is given by $(\kappa\cos\Phi)^2+z^2=1$,
$\kappa=2C/3\beta\Psi_{max}^2$, Eq. (35). Lines 1 - 8 correspond,
respectively, to $k=0.5$, 0.8, 0.95, 1, 1.05, 1.25, 2, and 10.
Lines 1 -3 describe the self-trapped mode, line 4 describes the
separatrix, lines 5 - 8 describe the wandering breather or new
tunneling mode of two weakly linked BECs. } \label{fig1}
\end{figure}

\begin{figure}[p]
\begin{center}
\includegraphics[angle=0, width=1\textwidth]{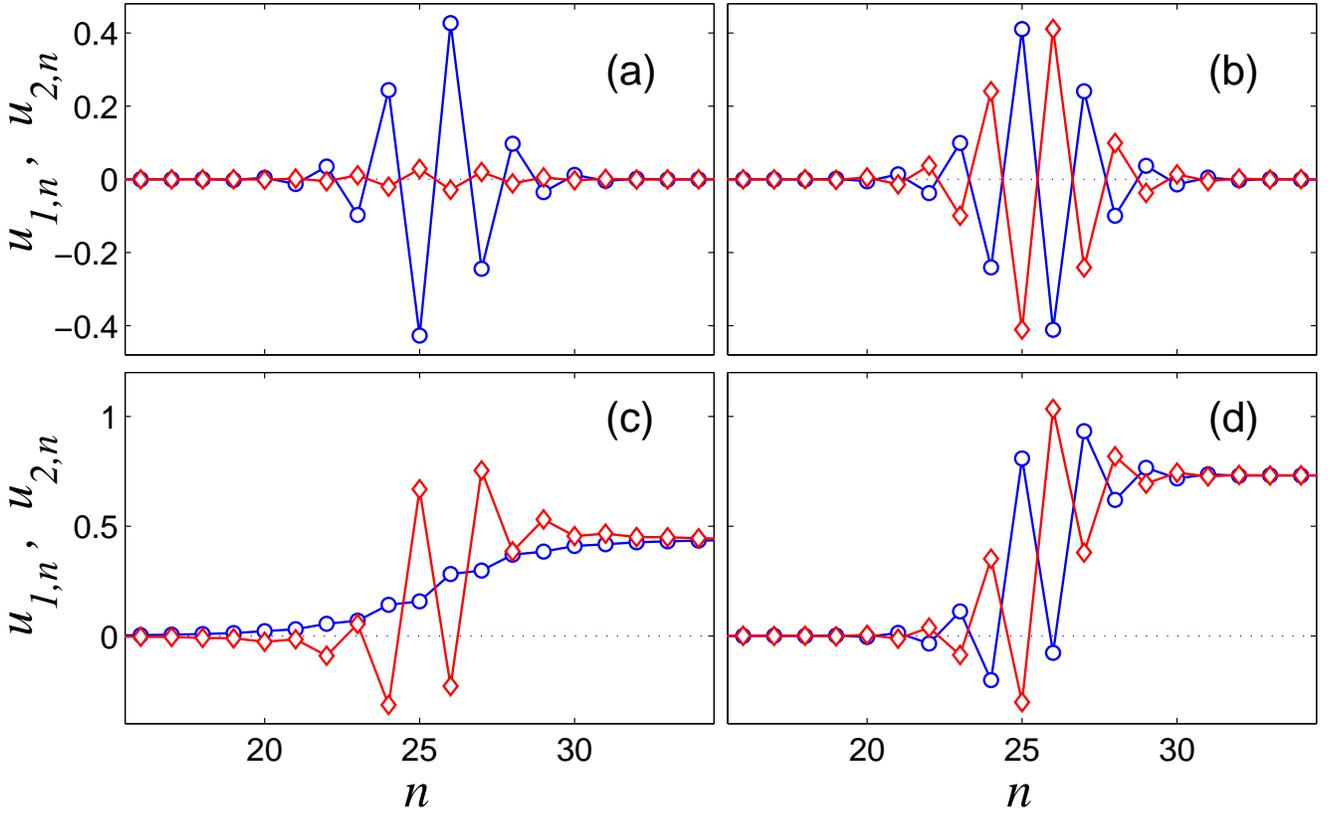}
\end{center}
\caption{\label{Fig2}\protect (color online) Displacement patterns
for (a) one-chain breather and (b) two-chain antisymmetric
breather with frequency $\omega=2.3$ for the quartic intra-chain
anharmonic potential ($\alpha=0$, $\beta=1$); (c) and (d) show
displacement patterns for one-chain and two-chain breathers,
respectively, in the presence of cubic and quartic intra-chain
anharmonic potentials ($\alpha=-1$, $\beta=1$) for the same
breather frequency. The patterns are shown at the moments when all
particle velocities are zero.}
\end{figure}

 \begin{figure}[p]
\begin{center}
\includegraphics[angle=0, width=1\textwidth]{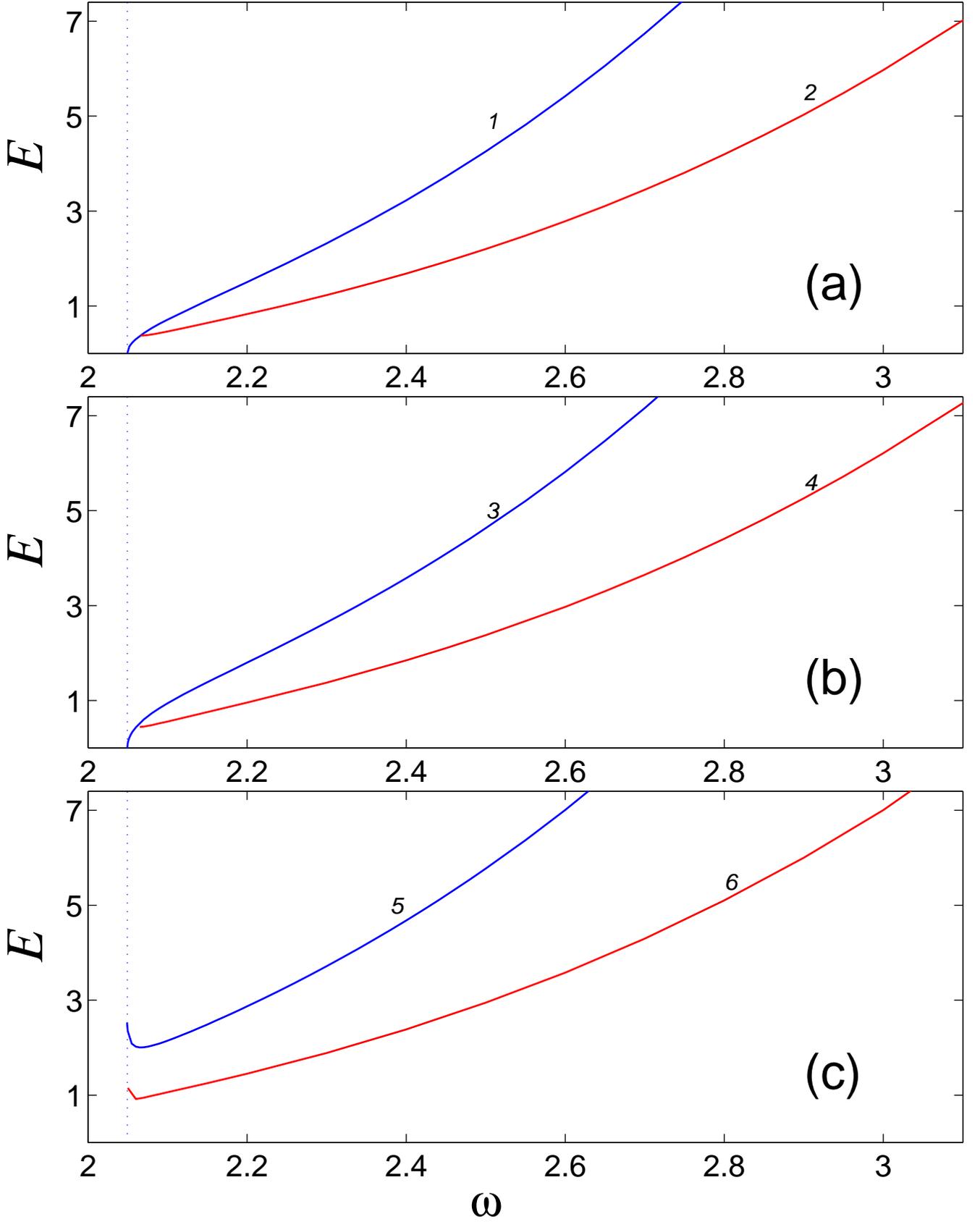}
\end{center}
\caption{\label{Fig3}\protect (color online) Dependence of
breather energy $E$ in a system of two coupled chains on breather
frequency $\omega$ for different strengths of cubic intra-chain
anharmonic potential: $\alpha=0$ (a), $\alpha=-0.5$ (b) and
$\alpha=-1$ (c), with the same strength of quartic anharmonic
potential $\beta=1$. Lines 1, 3, 5 correspond to antisymmetric
two-chain breather, lines 2,4, 6 correspond to one-chain breather.
All the lines correspond to the breathers with centers of symmetry
located in the middle of two nearest particles, cf. Figs. 2(a) -
2(d).  The dotted vertical line in all the plots shows the upper
boundary of phonon spectrum in the two-chain system
$\sqrt{4+2C}\approx 2.05$.}
\end{figure}

\begin{figure}[p]
\begin{center}
\includegraphics[angle=0, width=1\textwidth]{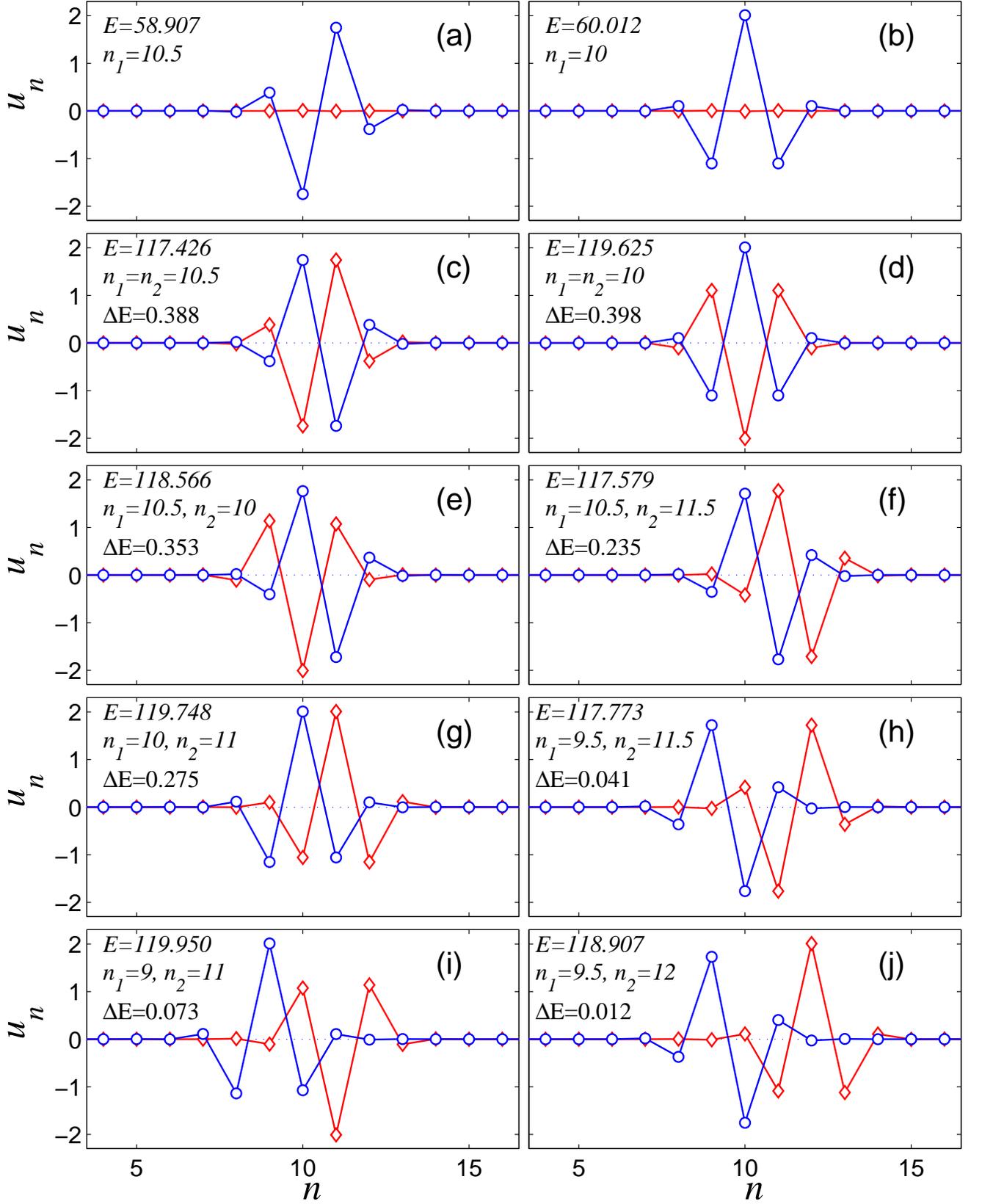}
\end{center}
\caption{\label{Fig4}\protect (color online) Displacement patterns
for two-chain breathers with $\omega=5$ in a system of two chains
with pure quartic intra-chain anharmonic potetial, with
$\alpha=0$, $\beta=1$. The patterns are shown at the moments when
all particle velocities are zero. Here $E$ is breather energy,
$n_{1}$ and $n_{1}$ are the locations of the centre of symmetry of
the breather in the $i$-th chian, $\Delta E=
E_{1}^{(1ch)}+E_{2}^{(1ch)} -E$ is the gain in energy due to
breather coupling (breather "binding  energy").}
\end{figure}

\begin{figure}[p]
\begin{center}
\includegraphics[angle=0, width=1\textwidth]{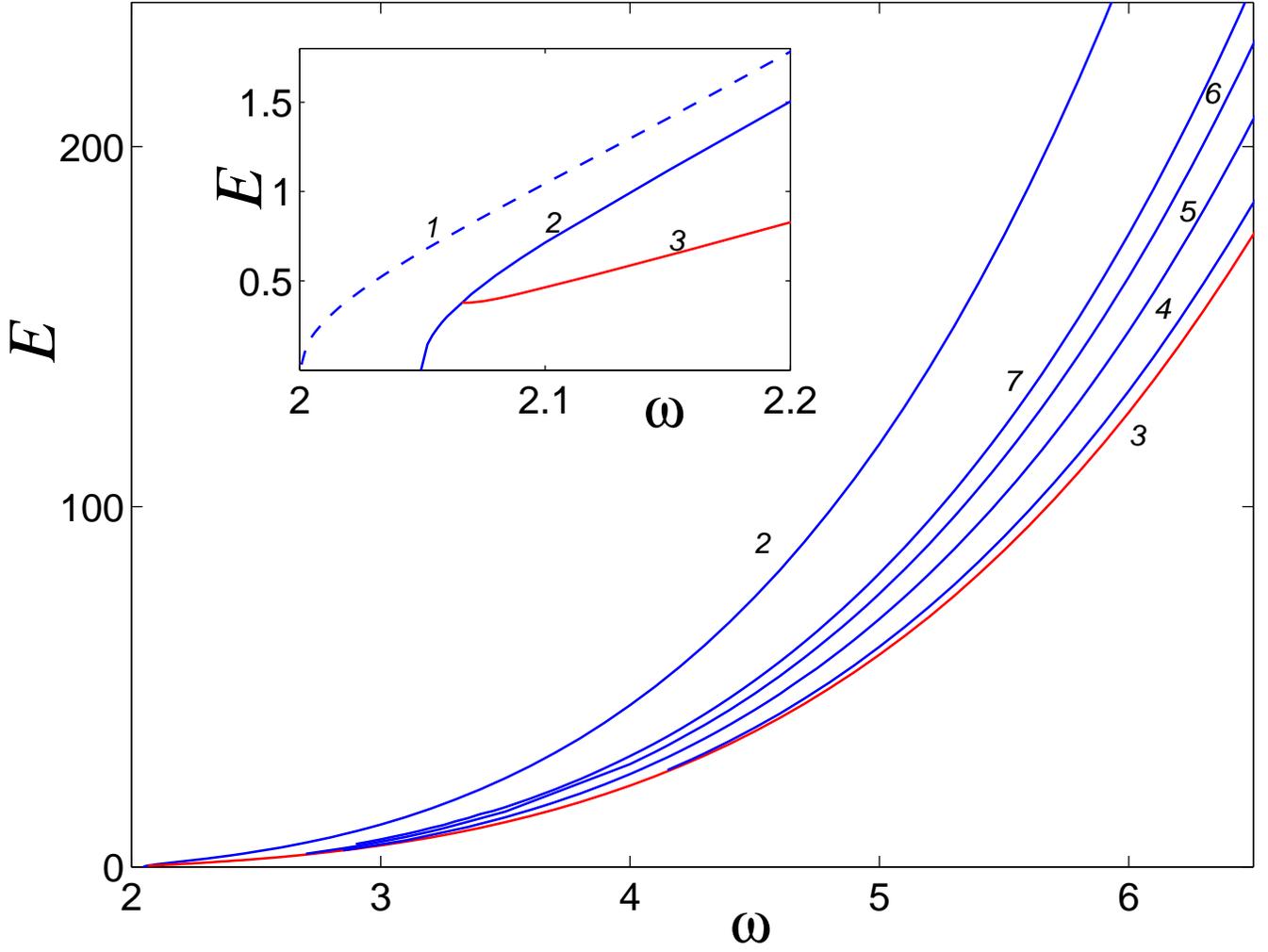}
\end{center}
\caption{\label{Fig5}\protect (color online) Breather energy $E$
in a system of two chains versus frequency $\omega$. Dashed line 1
in the inset corresponds to symmetric two-chain breather, line 2
describes two-chain antisymmetric breather, line 3 describes
one-chain breather. Lines 4 - 7 in the main plot  correspond to
two-chain excitations which couple breathers with two different
commensurate frequencies with ratios  $\omega_1:\omega_2=1:2$,
$\omega_1:\omega_2=2:3$, $\omega_1:\omega_2=3:4$
$\omega_1:\omega_2=4:5$, respectively. Frequency $\omega$ in the
plot always denotes $\omega_2$.}
\end{figure}

\begin{figure}[p]
\begin{center}
\includegraphics[angle=0, width=1\textwidth]{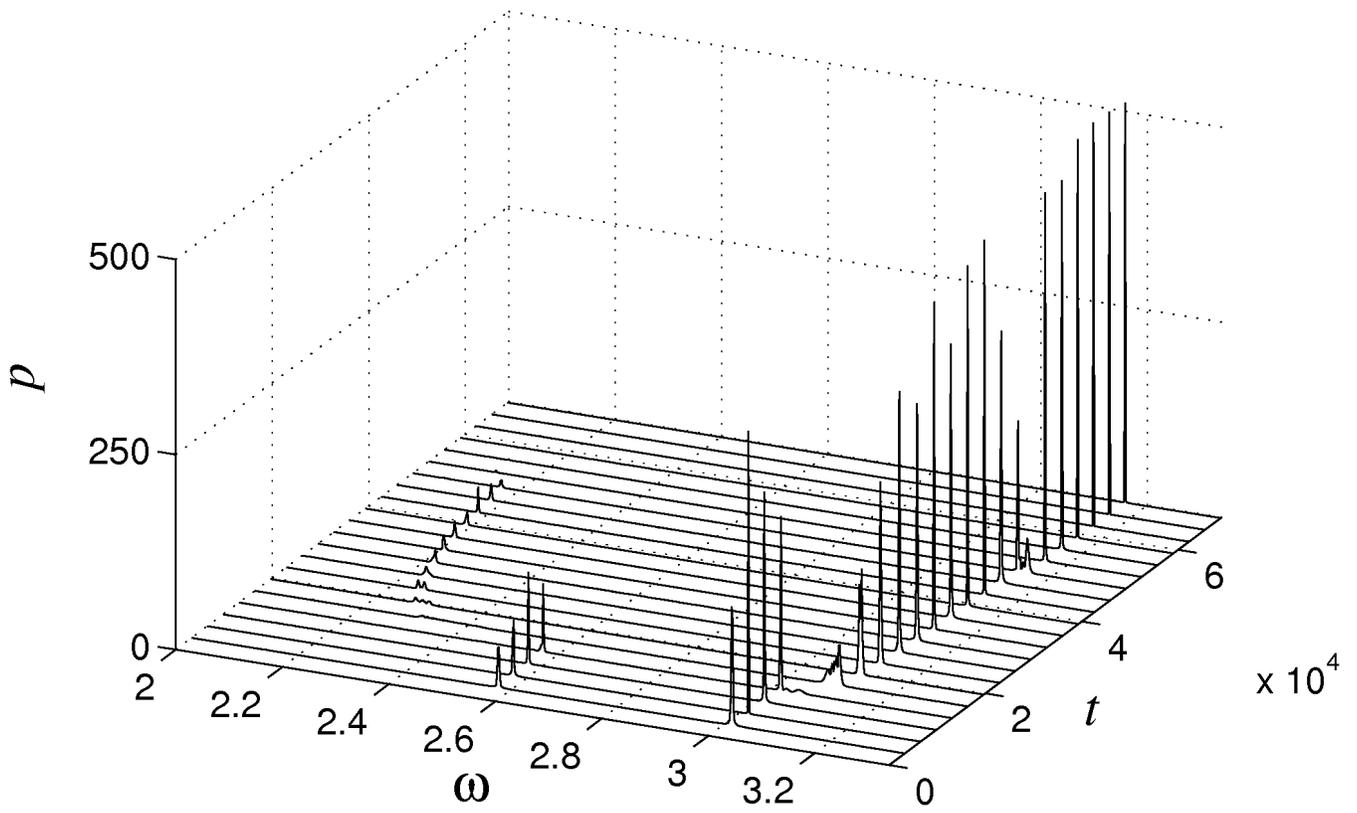}
\end{center}
\caption{\label{Fig6}\protect Time evolution of spectral density
$p(\omega)$ of nonlinear localized excitation of two weakly
coupled chains under the initial excitation of a breather with
frequency  $\omega_1=2.6$ in one chain and breather with frequency
$\omega_2=3$ in another chain.}
\end{figure}

\begin{figure}[p]
\begin{center}
\includegraphics[angle=0, width=0.7\textwidth]{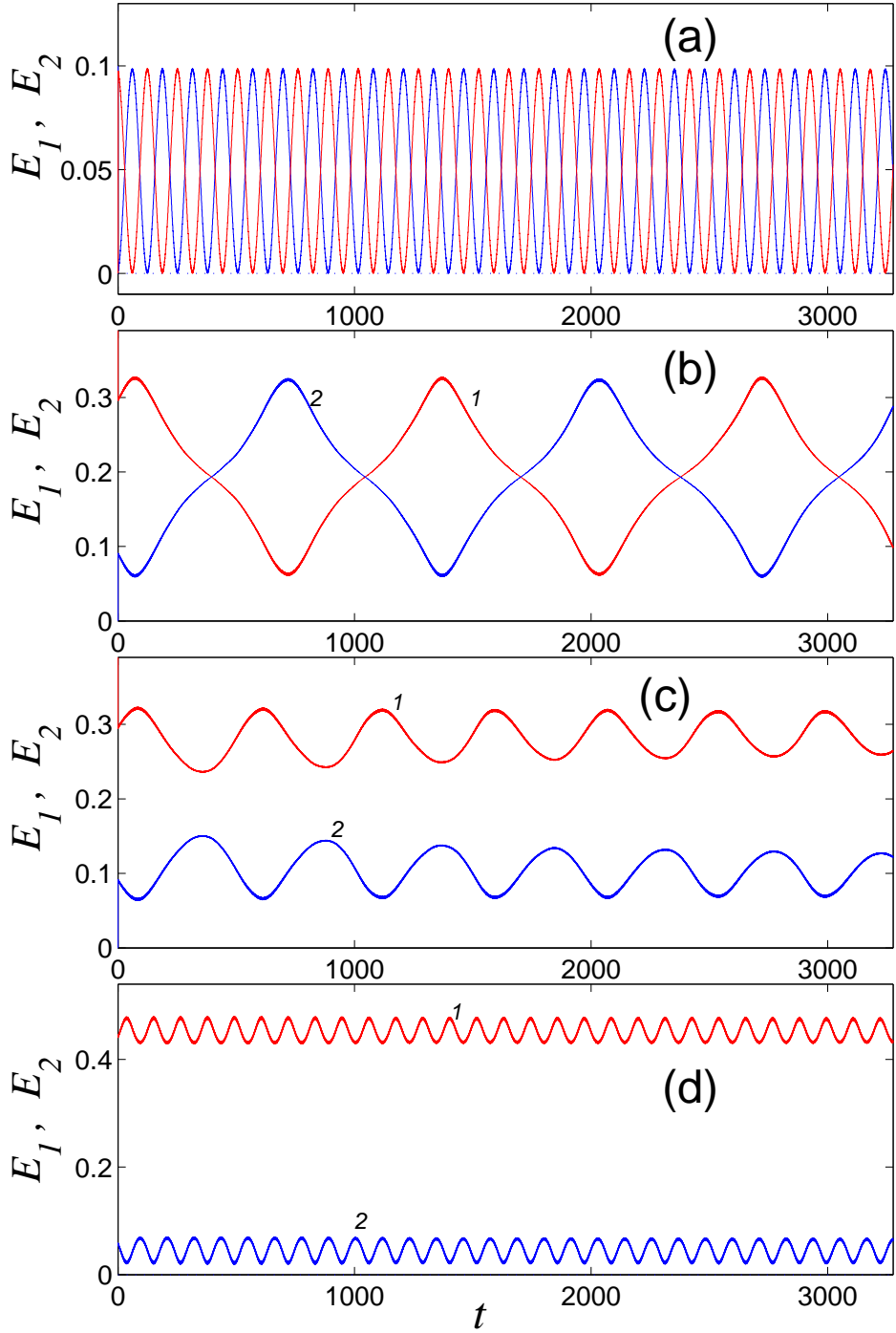}
\end{center}
\caption{\label{Fig7}\protect (color online) Time evolution of
chain energies $E_1$ (red) and $E_2$ (blue) of immovable breather
in chains 1 and 2 versus time $t$, obtained from numerical
solution of Eqs. (53) with the initial breather excitation in
chain 1 (for immovable chain 2) with frequency (a)
$\omega=2.0300$, (b) $\omega=2.0980$, (c) $\omega=2.0981$, and (d)
$\omega=2.1106$. Two identical chains with $\alpha=0$, $\beta=1$,
$C=0.1$ and absorbing edges were used in simulations.  Separatrix
solution of Eq. (39) corresponds to  $\omega=2.0976$. Time
evolution is shown after $t=10^5$ from the excitation instant when
all transients died out. }
\end{figure}

\begin{figure}[p]
\begin{center}
\includegraphics[angle=0, width=1\textwidth]{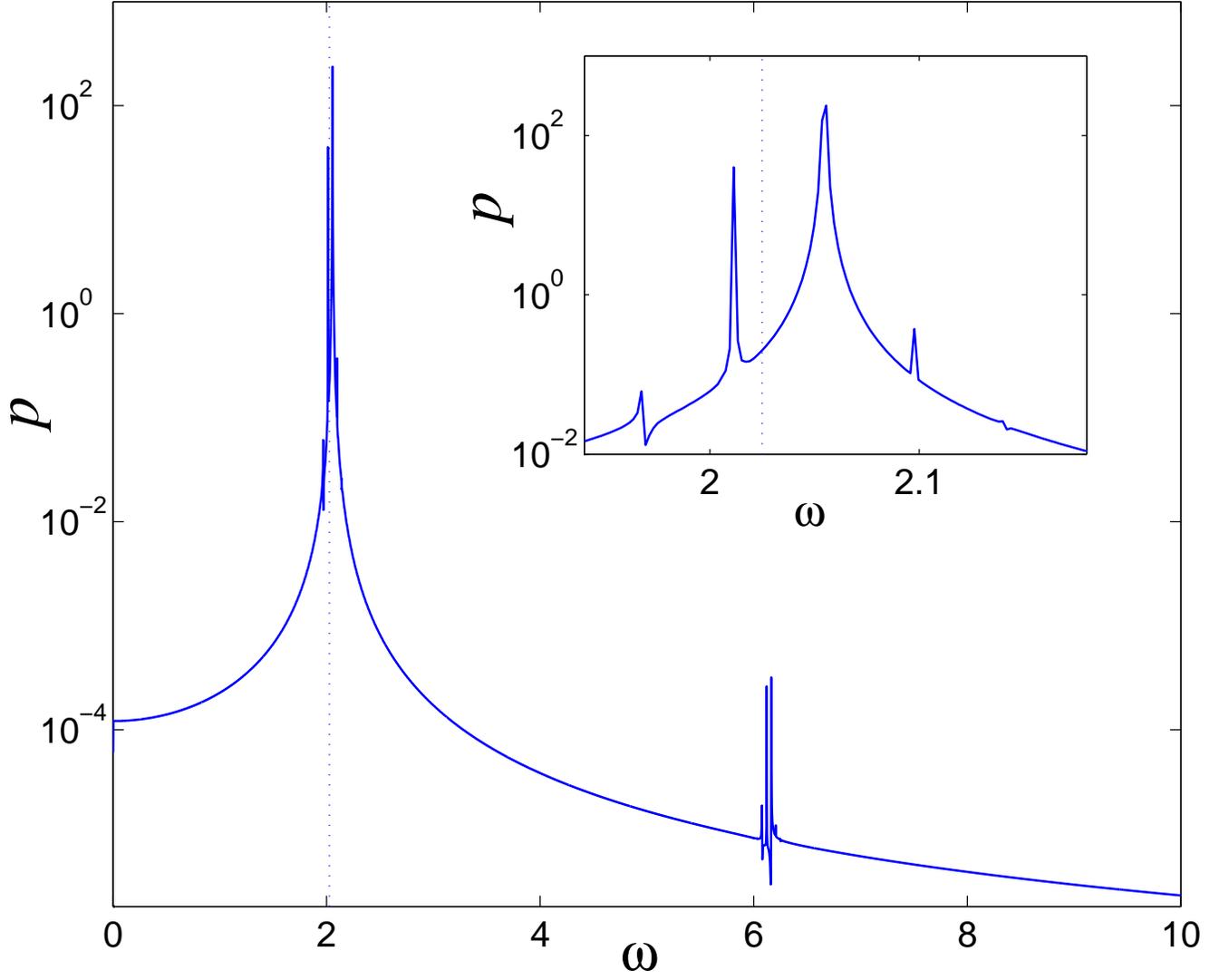}
\end{center}
\caption{\label{Fig8}\protect Spectral density $p(\omega)$ of
wandering breather in a system of two weakly coupled chains under
the initial excitation of a breather with frequency $\omega=2.05$.
Dotted vertical line indicates the upper bound of phonon frequency
$\omega_m =\sqrt{4+C}\approx 2.025$ in Eqs. (14). Inset shows fine
details of the spectral density in close vicinity of the
fundamental breather frequency. }
\end{figure}

\begin{figure}[p]
\begin{center}
\includegraphics[angle=0, width=1\textwidth]{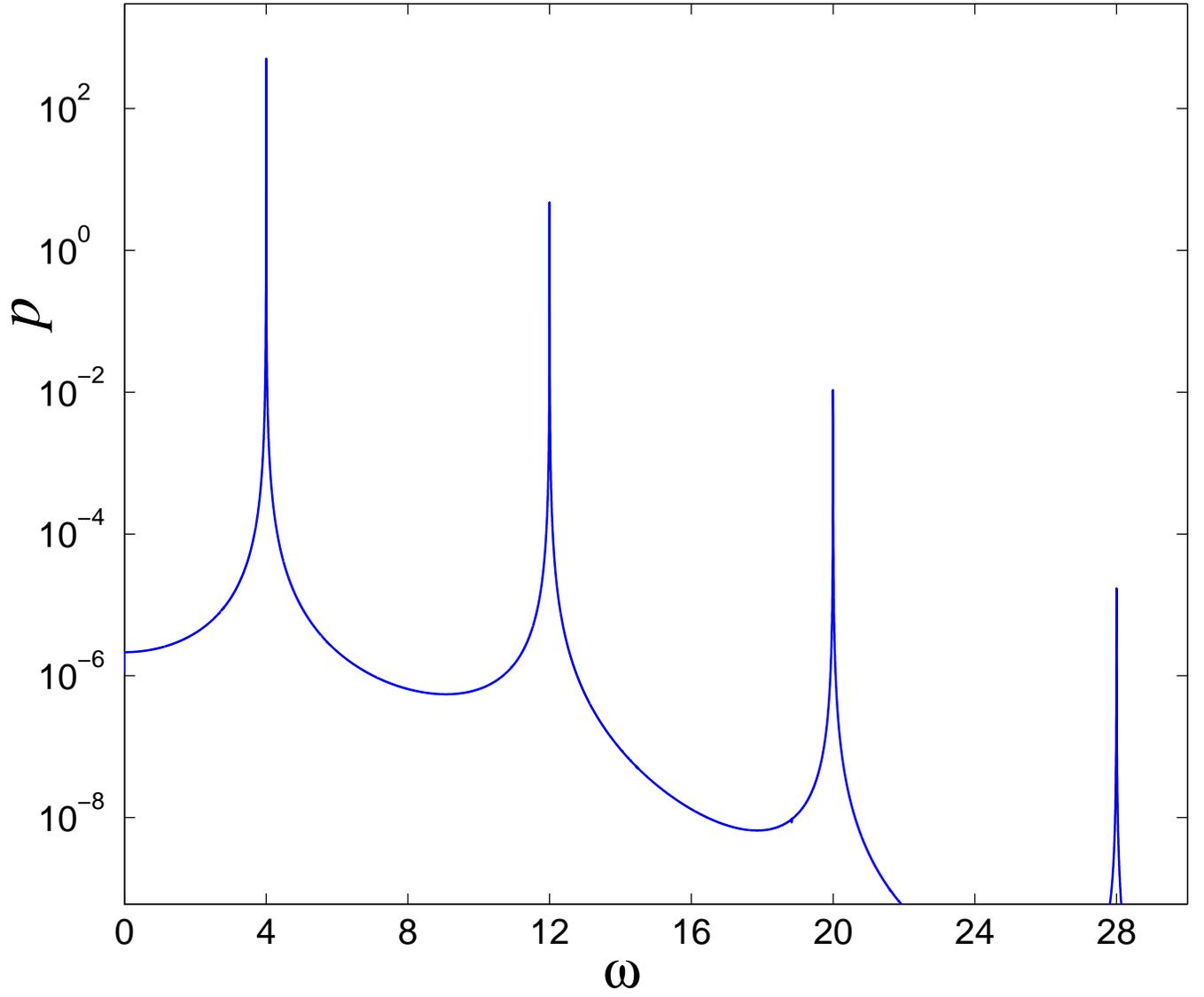}
\end{center}
\caption{\label{Fig9}\protect Spectral density $p(\omega)$ of
one-chain large-amplitude breather with frequency $\omega=4$ (in
self-trapped regime), in two coupled chains with $C=0.1$,
$\alpha=0$, $\beta=1$.}
\end{figure}

\begin{figure}[p]
\begin{center}
\includegraphics[angle=0, width=1\textwidth]{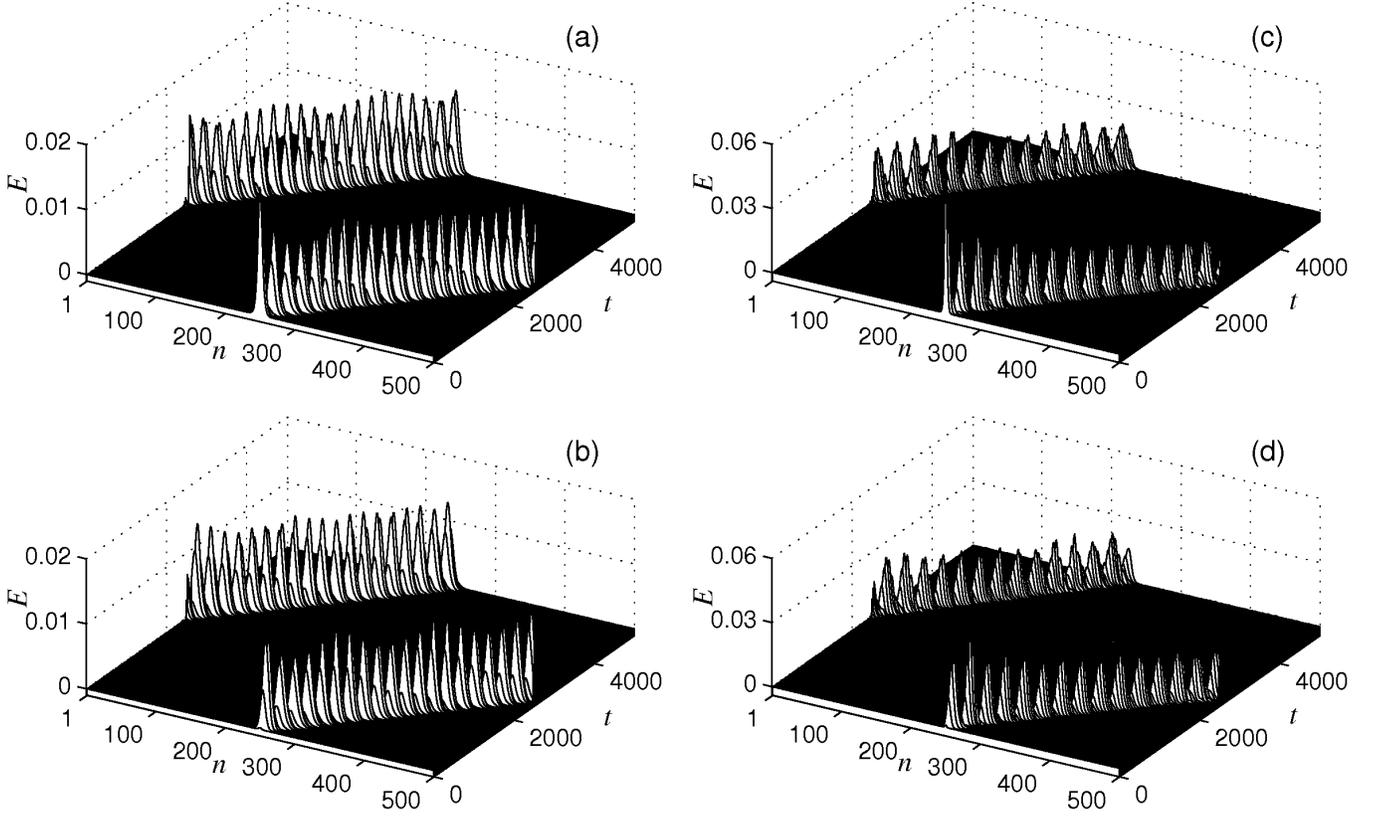}
\end{center}
\caption{\label{Fig10}\protect Periodic translation of
slowly-moving breather between  two coupled chains with
 $C=0.1$, $\alpha=0$, $\beta=1$. Time dependence of breather energy distribution
in chain 1, (a) and (c), and chain 2, (b) and (d), is shown. The
breather was initially excited in chain 1 with immovable chain 2,
with velocity $V=0.1$ and amplitude $\Psi_{max}=0.1$, (a) and (b),
and with velocity $V=0.1$ and amplitude $\Psi_{max}=0.2$, (c) and
(d). }
\end{figure}

 \begin{figure}[p]
\begin{center}
\includegraphics[angle=0, width=1\textwidth]{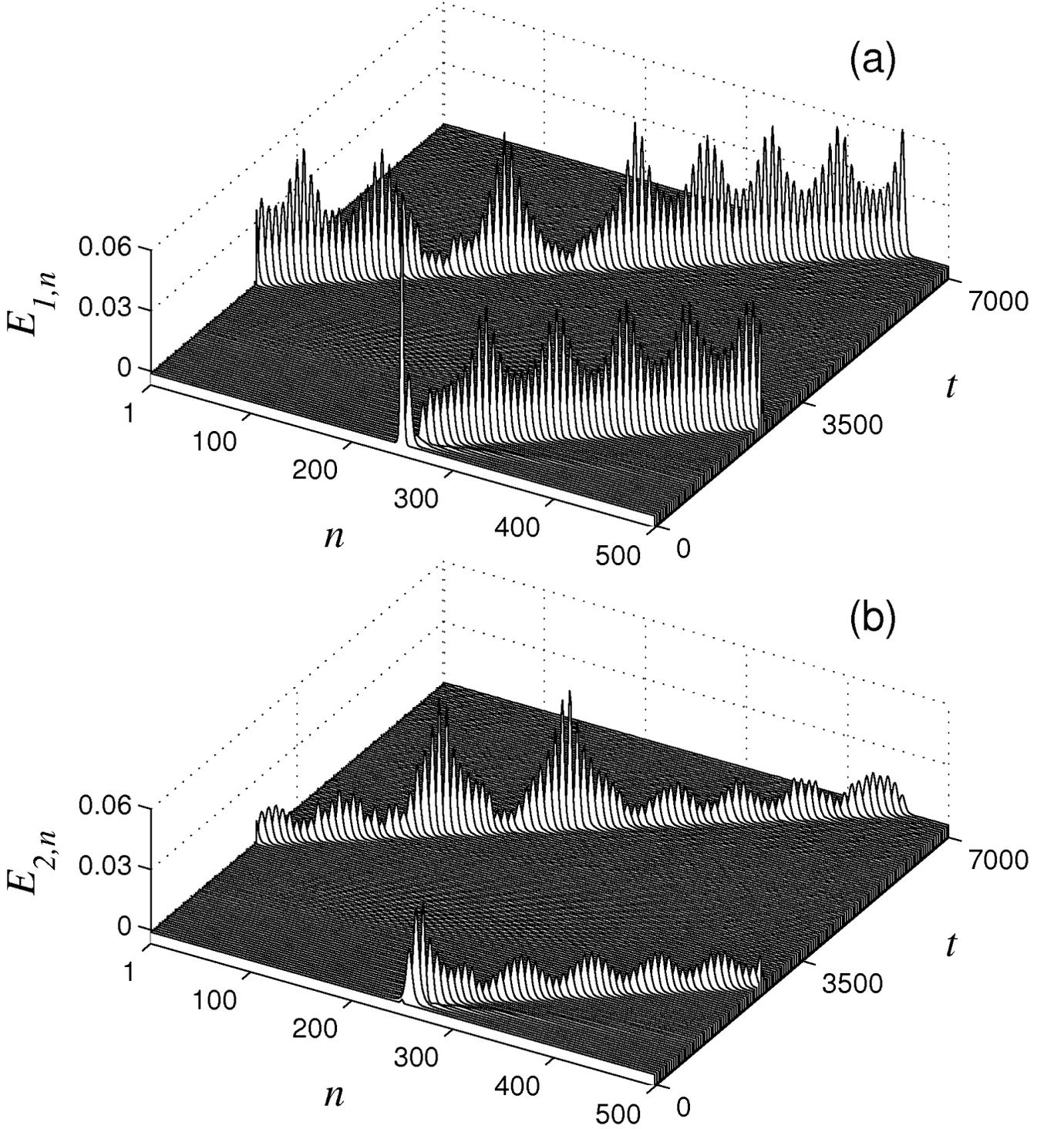}
\end{center}
\caption{\label{Fig11}\protect Energy of slowly-moving wandering
breather close to the separatrix in  chain 1 (a) and chain 2 (b)
versus time and site. The breather is initially excited in chain 1
with immovable chain 2, with velocity $V$$=$0.1 and amplitude
$\Psi_{max}$$=$0.26. The separatrix solution of Eq.\,(\ref{e.27})
corresponds to $\Psi_{max}$=0.2582. Separatrix-like dynamics,
similar to the one shown in Fig. 7(b), is well established for
$t$$\geq$4000.}
\end{figure}

\begin{figure}[p]
\begin{center}
\includegraphics[angle=0, width=1\textwidth]{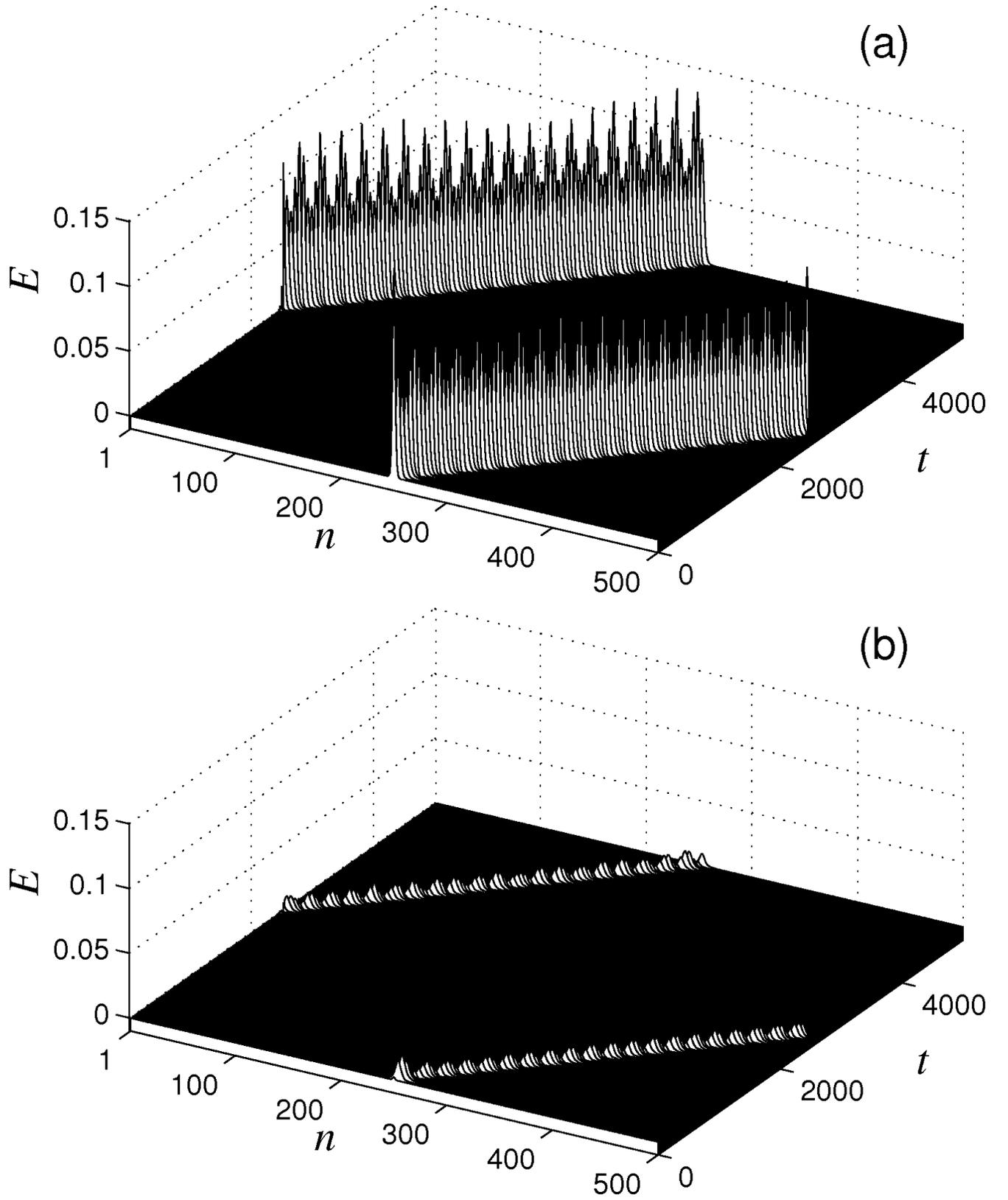}
\end{center}
\caption{\label{Fig12}\protect Distribution of breather energy in
chain 1 (a) and chain 2 (b) versus time in the self-trapped regime
of slowly-moving breather in two coupled chains with $C=0.1$,
$\alpha=0$, $\beta=1$. The breather was initially excited in chain
1 with immovable chain 2, with velocity $V=0.1$ and amplitude
$\Psi_{max}=0.3$.}
\end{figure}

\begin{figure}[p]
\begin{center}
\includegraphics[angle=0, width=1\textwidth]{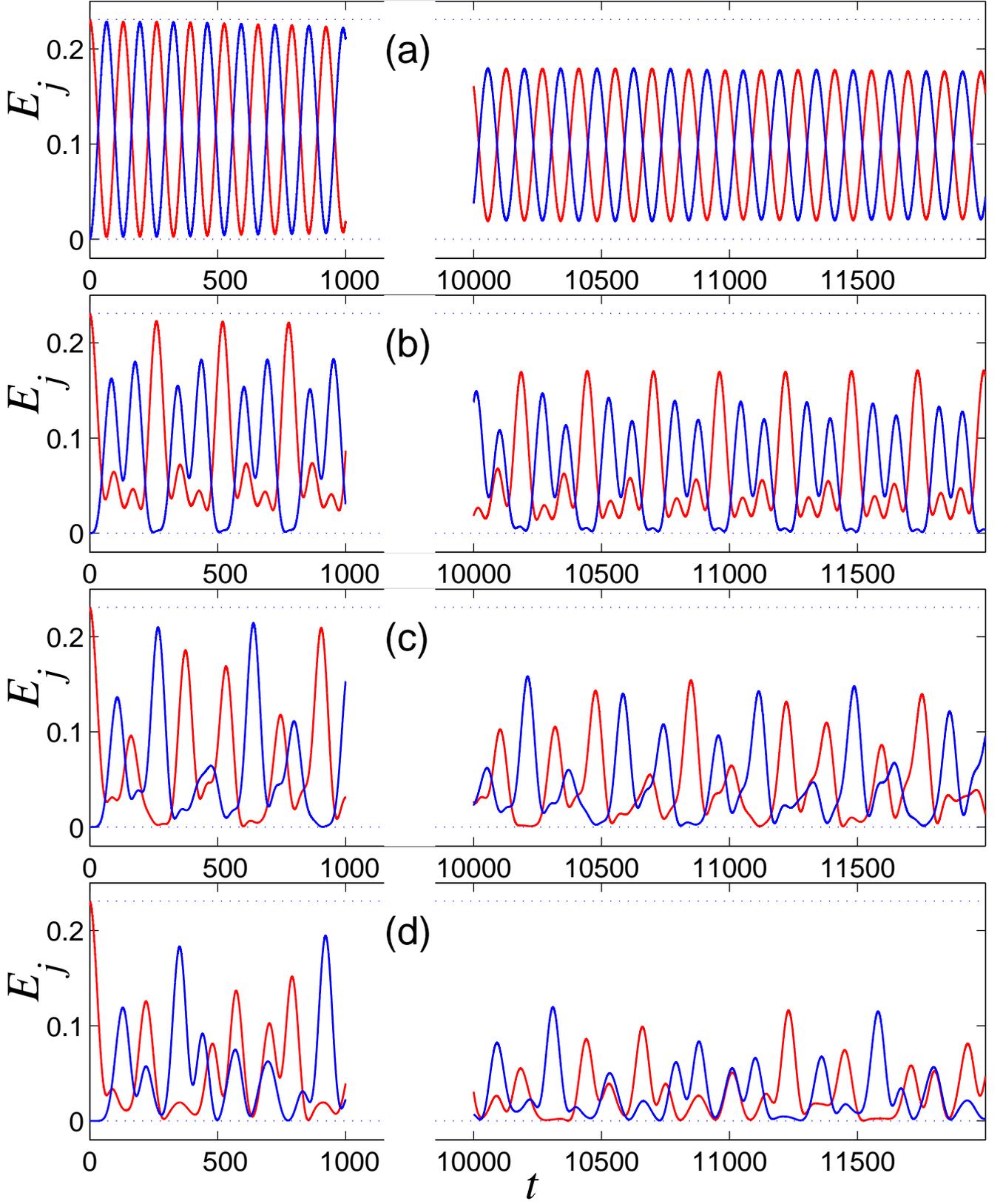}
\end{center}
\caption{\label{Fig13}\protect (color online) Time dependence of
breather energy in the first (red) and last (blue) chains in a
system of 2 (a), 3 (b), 4 (c), and 5 (d) chains with $C=0.1$,
$\alpha=0$, $\beta=1$. The breather is always initially excited
with frequency $\omega=2.05$ in chain 1, with immovable rest of
the chains.}
\end{figure}

\begin{figure}[p]
\begin{center}
\includegraphics[angle=0, width=1\textwidth]{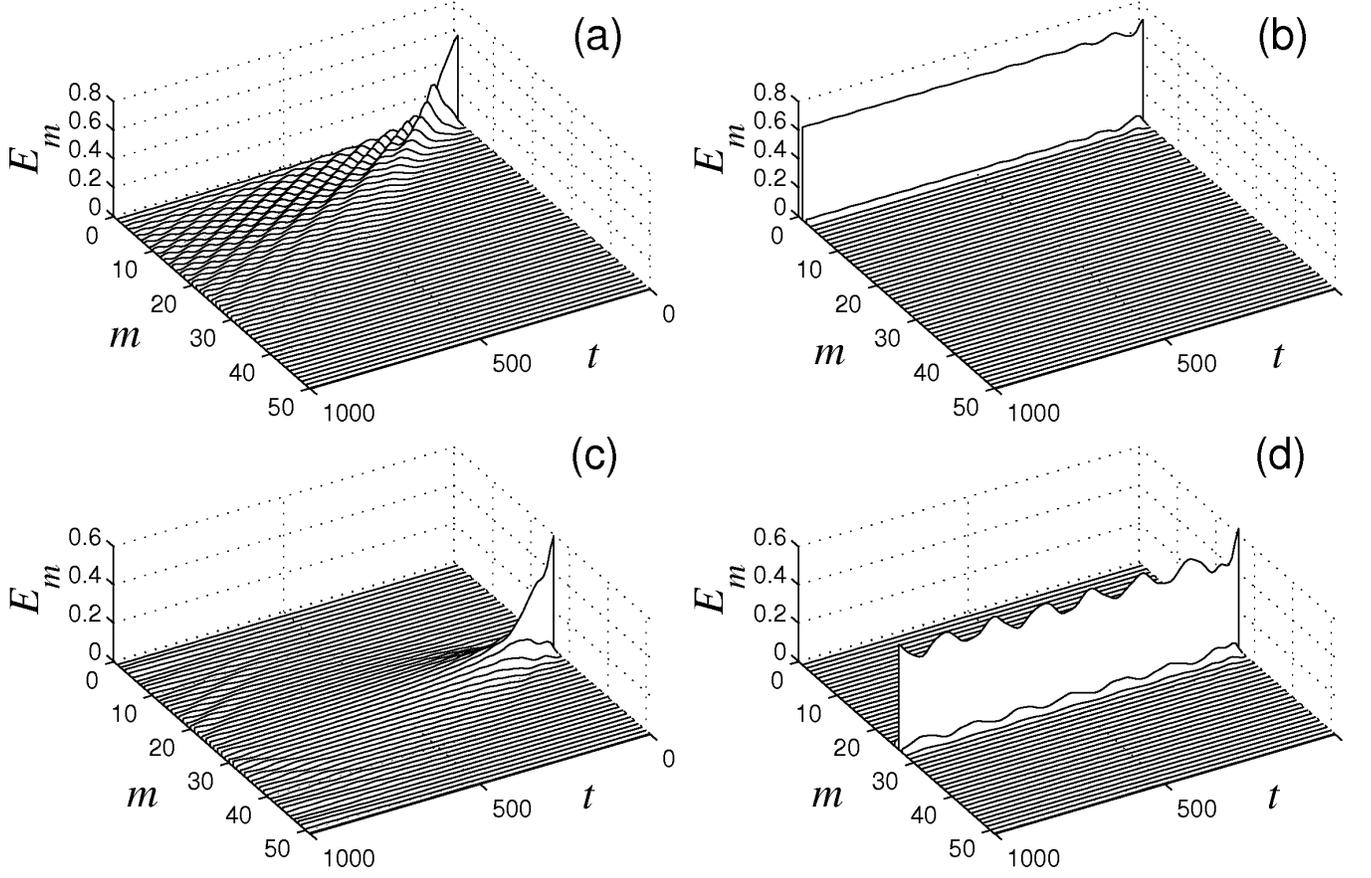}
\end{center}
\caption{\label{Fig14}\protect Time dependence of breather energy
distribution between chains in a system of 50 coupled chains with
 $C=0.1$, $\alpha=0$, $\beta=1$. Here $m$ is chain number and
 $E_m$ is energy of the $m$-th chain.
The breather was initially excited with frequency $\omega=2.14$ in
chain 1 (a), with frequency $\omega=2.15$ in chain 1 (b), with
frequency $\omega=2.16$ in chain 25 (c), with frequency
$\omega=2.17$ in chain 25 (d), with immovable rest of chains. }
\end{figure}

\begin{figure}[p]
\begin{center}
\includegraphics[angle=0, width=1\textwidth]{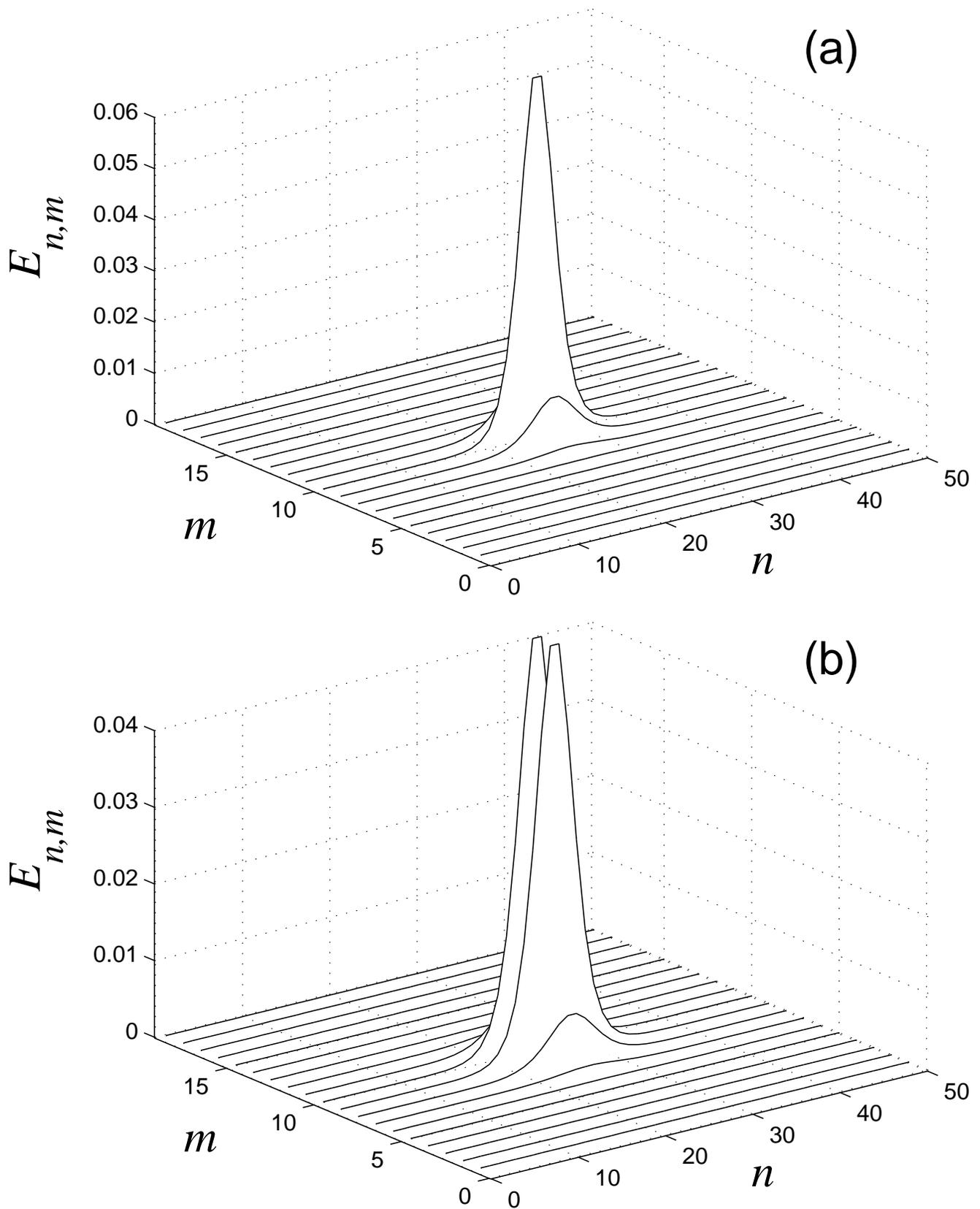}
\end{center}
\caption{\label{Fig15}\protect  Distribution of vibration energy
in 2D lattice of weakly coupled nonlinear chains, with $\alpha=0$,
$\beta=1$ and $C=0.1$, along (n) and across (m) the chains  for
one-chain (a) and anti-phase two-chain (b) breather with frequency
$\omega=2.12$. }
\end{figure}

\begin{figure}[p]
\begin{center}
\includegraphics[angle=0, width=1\textwidth]{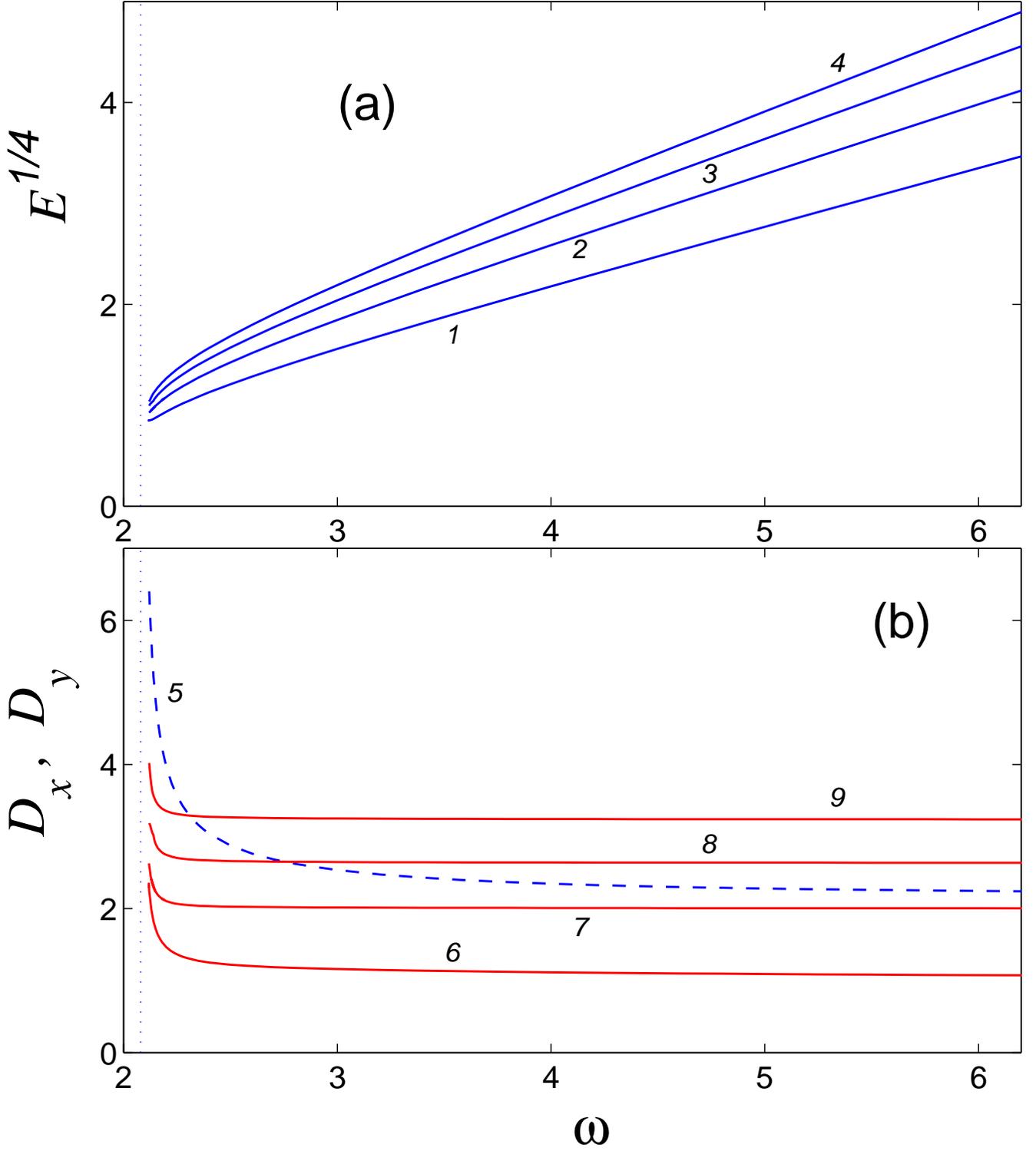}
\end{center}
\caption{\label{Fig16}\protect Dependence of breather vibrational
energy (a),  and longitudinal $D_x$ and transverse $D_y$ breather
diameters (b) versus breather frequency for one-chain (curves
1,5,6), anti-phase two-chain (curves 2,5,7), anti-phase
three-chain (curves 3,5,8), and anti-phase four-chain (curves
4,5,9) breathers in a multi-chain system ($M=20$, $N=50$). Dashed
line corresponds to the longitudinal breather diameter. Vertical
dotted line shows the cut off phonon frequency.}
\end{figure}

\end{document}